\documentclass[a4paper,11pt]{article}

\usepackage{rotating}
\usepackage[english]{babel}
\usepackage{graphicx}
\usepackage{subfig}
\usepackage{float}
\usepackage{amsmath}
\usepackage{amssymb}
\usepackage{amsthm}
\usepackage{multirow}
\usepackage{hyperref}
\usepackage{mciteplus}
\usepackage{verbatim}
\usepackage{cancel}
\usepackage{lscape}
\usepackage{diagbox}
\usepackage[usenames, dvipsnames]{xcolor}

\def\gtwid{\mathrel{\raise.3ex\hbox{$>$\kern-.75em\lower1ex\hbox{$\sim
$}}}}
\def\vio{\mathrel{\hbox{$E$\kern-.60em\hbox{$/
$}}}}
\def\gev{{\rm GeV}}
\def\beq{\begin{equation}}
\def\eeq{\end{equation}}
\def\bea{\begin{eqnarray}}
\def\eea{\end{eqnarray}}

\def\h1{\ensuremath{h_1}}
\def\h2{\ensuremath{h_2}}
\def\A{\ensuremath{A}}
\def\tanb{\ensuremath{\tan\beta}}
\def\half{\ensuremath{\frac{1}{2}}}

\newcommand{\hobs}{H_{\rm obs}}
\newcommand{\hp}{H^+}
\newcommand{\hm}{H^-}
\newcommand{\hpm}{H^\pm}
\newcommand{\hpb}{\textit{\textbf{H}}$^+$}
\newcommand{\hmb}{\textit{\textbf{H}}$^-$}
\newcommand{\hpmb}{\textit{\textbf{H}}$^\pm$}

\newcommand{\wmi}{W^-}
\newcommand{\wpm}{W^\pm}

\newcommand{\wpmb}{\textit{\textbf{W}}$^\pm$}

\newcommand{\cm}{\checkmark}

\newcommand{\rbtcb}[1]{\fcolorbox{black}{White}{#1}}
\newcommand{\bbtcr}[1]{\fcolorbox{black}{White}{#1}}
\newcommand{\mr}[1]{\multirow{#1}{*}}
\newcommand{\boldit}[1]{\textit{\textbf{#1}}}

\begin{document}

\thispagestyle{empty}
\begin{flushright}
KIAS-P18109\\
\end{flushright}
\vspace*{2.0cm}

\begin{center}
{\Large \bf Electroweak production of  multiple (pseudo)scalars in the 2HDM}\\ 
\vspace{.3in}
{\large Rikard Enberg$^a$, William Klemm$^{a,b}$, Stefano
  Moretti$^c$ and Shoaib Munir$^{d,e}$} \\[0.25cm]
{\sl $^a$ Department of Physics and Astronomy, \\ 
Uppsala University, Box 516, SE-751 20 Uppsala, Sweden.}\\[0.25cm]
{\sl $^b$School of Physics \& Astronomy,\\
 University of Manchester, Manchester M13 9PL, UK.}\\[0.25cm]
{\sl $^{c}$ School of Physics \& Astronomy, \\
University of Southampton, Southampton SO17 1BJ, UK.} \\[0.25cm]
{\sl $^{d}$ School of Physics, Korea Institute for Advanced Study,\\ 
Seoul 130-722, Republic of Korea.} \\[0.25cm]
{\sl $^{e}$ East African Institute for Fundamental Research (ICTP-EAIFR), \\
University of Rwanda, Kigali, Rwanda.} \\[0.25cm]
\end{center}
\vspace{0.3in}

\begin{abstract} 
The two-Higgs Doublet Model (2HDM) is the most minimal extension of the Standard Model (SM) containing extra Higgs doublet fields. Given the multiplicity of Higgs states in a 2HDM, its Higgs potential is significantly more involved than the SM one. Importantly, it contains a multitude of Higgs triple self-couplings, unlike the SM, which only has one. These interactions are key to understanding the phenomenology of the 2HDM, as they uniquely determine the form of the potential. Several studies analysing the prospects of measuring these couplings at the Large Hadron Collider (LHC) have found them to be quite low generally. However, such studies have largely concentrated on Higgs pair-production induced by gluon-gluon scattering, either via direct annihilation or followed by their splitting into $b$-(anti)quark pairs, which in turn annihilate leaving behind spectator $b$-(anti)quarks. Both of these channels are therefore governed by QCD dynamics. We compare here the yields of such channels to those initiated by (primarily) valence quarks, which involve Electro-Weak (EW) interactions only, for neutral multi-Higgs final states. We find that EW production can be dominant over QCD production for certain final state combinations. We also illustrate that charged final states, which can only be produced via EW modes, could serve as important probes of some $H^\pm$ triple couplings, that are inaccessible in QCD-induced processes, during Run 2 and 3 of the LHC. Our analysis covers regions of the parameter space of the Type-I 2HDM that have escaped the most up-to-date experimental constraints coming from EW precision data, LHC measurements of the 125\,GeV Higgs boson properties, searches for additional Higgs states, and flavour physics.  
\\[0.5cm]
\noindent
{\sl We dedicate this work to the memory of Prof. W. James Stirling, an example to never forget.}
\end{abstract}
\vskip 10mm
\noindent {\footnotesize $^\dagger$E-mails:\\
{\tt 
{rikard.enberg@physics.uu.se},
{william.klemm@physics.uu.se},
{s.moretti@soton.ac.uk},
{smunir@eaifr.org}
}
}

\newpage
\section{\label{intro}Introduction}

The 2012 discovery of a neutral Higgs boson \cite{Aad:2012tfa,Chatrchyan:2012xdj}, $\hobs$, with a mass near 125~\gev, is strong evidence for gauge boson masses being induced by the Higgs mechanism of Electroweak Symmetry Breaking (EWSB). While the Higgs boson data collected at the LHC is still consistent with the minimal EWSB dynamics of the SM, some other experimental results cannot be reconciled with it. In particular, certain anomalies in the flavour sector \cite{Bennett:2006fi,Aaij:2014ora,Aaij:2017vbb,Aaij:2015yra,Aaij:2017deq} are far more compatible with an extended Higgs sector \cite{Cherchiglia:2016eui,Crivellin:2012ye,Celis:2012dk,Sakaki:2013bfa,Bauer:2015knc} than with the SM. In view of this, as the $\hobs$ state emerges from a Higgs doublet in the SM, the phenomenology of its minimal extension by another Higgs doublet, which results in the two-Higgs Doublet Model, deserves particular attention.

In the 2HDM Higgs sector, five physical states emerge after EWSB: three neutral, of which two are scalars ($h$ and $H$, with $m_h<m_H$) and one a pseudoscalar ($A$), plus a charged pair ($\hpm$). The theory of this scenario is well-understood (see, e.g., \cite{Branco:2011iw,Gunion:1989we}), but its phenomenological investigation is far from complete at present. In particular, while there exist some indications of what the accessible discovery channels of the additional Higgs bosons of a 2HDM could be at the LHC, little effort has been spent on assessing which are the most suitable channels to pin down the specific nature of the underlying Higgs dynamics. The reason is that there are several incarnations of the 2HDM and, although each of them yields a different phenomenological pattern in general, there exists a significant level of degeneracy among them if only the production and decay channels of a single Higgs state are studied. Indeed, for an unequivocal extraction of a 2HDM scenario involved in EWSB, the various components of the scalar potential ought to be accessed experimentally. This makes the study of multi-Higgs final states mandatory.

In the context of the LHC, several analyses exist in literature, addressing double, or even triple, Higgs production, assuming a 2HDM to be the underlying framework (see, e.g., Ref. \cite{Arhrib:2009hc,Hespel:2014sla} for a review). However, the majority of such analyses have concentrated on production modes induced by QCD dynamics, notably gluon-gluon ($gg$) fusion into a (neutral) pair of Higgs states. These pairs emerge either from a primary Higgs state (resonantly or otherwise) or as Higgs-strahlung from a box diagram involving heavy fermion loops. Alternatively, because Higgs couplings to quarks are of Yukawa type (i.e., proportional to the quark mass), the $b\bar b$ scattering channel has also been exploited. It should be noted that $b$-quarks are not valence partons and are therefore produced from a (double) gluon splitting. Hence this channel is also intrinsically $gg$-induced. 

While these QCD processes clearly afford one the possibility of the direct measurements of a number of terms in the 2HDM Lagrangian, the complete list of these terms is much longer. In order to remedy this, we study here double and triple Higgs boson production in $q\bar q^{(')}$-induced EW interactions, where $q$ represents predominantly a valence $u,d$, in the Type-I 2HDM. This theoretical scenario has been shown to yield spectacular signals involving light neutral Higgs states, with a mass smaller than that of $\hobs$, that are potentially accessible at the LHC, see Refs. \cite{Arhrib:2017uon,Enberg:2017gyo,Arhrib:2017wmo,Enberg:2016ygw}. Here, we assess the complementary portion of the Type-I 2HDM parameter space, wherein the lighter of the two scalar Higgs states has a mass of 125\,GeV, along the lines of \cite{Arhrib:2016wpw}, which considered a similar setup but concentrated exclusively on charged Higgs boson signals. We will argue that the cross sections for the production of some of these double (and triple) Higgs final states could be accessible within the already scheduled LHC Runs. We will in particular show that in certain cases not only can these cross sections be larger for EW processes compared to the QCD-initiated processes, but the former can also possibly provide access to some of the Higgs self-couplings that none of the latter can.

The article is organised as fellows. In Sec. \ref{sec:model} we review in some detail the various types of minimally flavour-violating 2HDM and identify the Higgs-Higgs and Higgs-gauge couplings available in it. In Sec. \ref{sec:scans} we discuss parameter space regions of the Type-I 2HDM which are amenable to LHC investigation in multi-Higgs final states, satisfying all the theoretical and experimental constraints of relevance. In Sec. \ref{sec:results} we discuss our results. Finally, we present our conclusions in Sec. \ref{sec:summa}.

\section{\label{sec:model} The two-Higgs Doublet Model}
The 2HDM contains two Higgs doublet fields, $\Phi_1$ and $\Phi_2$, and its most general potential can be written as
\begin{equation}
\begin{split}
\mathcal{V}_{\text{2HDM}}  &= m_{11}^2\Phi_1^\dagger\Phi_1+ m_{22}^2\Phi_2^\dagger\Phi_2
-[m_{12}^2\Phi_1^\dagger\Phi_2+ \, \text{h.c.} ] \\
& +\half\lambda_1(\Phi_1^\dagger\Phi_1)^2
+\half\lambda_2(\Phi_2^\dagger\Phi_2)^2
+\lambda_3(\Phi_1^\dagger\Phi_1)(\Phi_2^\dagger\Phi_2)
+\lambda_4(\Phi_1^\dagger\Phi_2)(\Phi_2^\dagger\Phi_1) \\
& +\left\{\half\lambda_5(\Phi_1^\dagger\Phi_2)^2
+\big[\lambda_6(\Phi_1^\dagger\Phi_1)
+\lambda_7(\Phi_2^\dagger\Phi_2)\big]
\Phi_1^\dagger\Phi_2+\, \text{h.c.}\right\}\,.
\label{eq:2hdmpot}
\end{split}
\end{equation}
Upon EWSB, $\Phi_1$ and $\Phi_2$ are defined in terms of their respective vacuum expectation values $v_1$ and $v_2$, the physical Higgs states $h$, $H$, $A$ and $H^\pm$, and the Goldstone bosons $G$ and $G^\pm$ as
\begin{equation}
\Phi_1=\frac{1}{\sqrt{2}}\left(\begin{array}{c}
\displaystyle \sqrt{2}\left(G^+\cos\beta -H^+\sin\beta\right)  \\
\displaystyle v_1-h\sin\alpha+H\cos\alpha+\mathrm{i}\left( G\cos\beta-A\sin\beta \right)
\end{array}
\right),
\end{equation}
\begin{equation}
\Phi_2=\frac{1}{\sqrt{2}}\left(\begin{array}{c}
\displaystyle \sqrt{2}\left(G^+\sin\beta +H^+\cos\beta\right)  \\
\displaystyle v_2+h\cos\alpha+H\sin\alpha+\mathrm{i}\left( G\sin\beta+A\cos\beta \right)
\end{array}
\right),
\end{equation}
where $\alpha$ is the mixing angle of the CP-even interaction states and $\tanb\equiv v_1/v_2$. Upon minimisation of the Higgs potential in Eq.\,(\ref{eq:2hdmpot}), after rewriting it in terms of these expanded fields, the bare masses $m_{11}^2$ and $m_{22}^2$ get replaced by $v_{1,2}$. Similarly, the quartic couplings $\lambda_{1-5}$ in Eq.\,(\ref{eq:2hdmpot}) can be traded for the masses of the four physical Higgs bosons as well as the mixing parameter $\sin(\beta-\alpha)$. The free parameters of a 2HDM thus include $m_h,\,m_H,\,m_A,\,m_{H^\pm},\lambda_6,\,\lambda_7,\,m_{12}^2,\,\tanb$ and $\sin(\beta-\alpha)$.

If all the SM fermions couple to both the Higgs fields of a 2HDM, it can lead to dangerous flavour-changing neutral currents (FCNCs). In order to avoid large FCNCs, the most general approach taken is to enforce a $Z_2$ symmetry on the Lagrangian, so that each type of fermion only couples to one of the doublets \cite{Glashow:1976nt,Paschos:1976ay}. This symmetry is softly broken by the $m_{12}^2$ term in the Higgs potential above and explicitly broken by the $\lambda_{6,7}$ terms. In the following we restrict ourselves to the CP-conserving case $\lambda_6=\lambda_7=0$.

The Type-I 2HDM is obtained if (conventionally) $\Phi_1 \to -\Phi_1$ under the $Z_2$ symmetry, so that all the quarks and charged leptons couple only to $\Phi_2$. On the other hand, the Type-II 2HDM observes the transformation property $\Phi_1 \to -\Phi_1,\, d_R^i\to -d_R^i,\,e_R^i\to -e_R^i$, so that only these mutually couple, while the up-type quarks couple instead to $\Phi_2$. The Type-III (or Type Y or ‘flipped’) model is built such that $\Phi_2$ couples to the up-type quarks and the leptons and $\Phi_1$ couples to the down-type quarks only while in the Type-IV (or Type X or ‘lepton-specific’) model $\Phi_2$ couples to all the quarks and $\Phi_1$ to all the leptons.   In this study, we will concentrate on the Type-I 2HDM, for whose allowed parameter space the relevance of the aforementioned EW processes with respect to the QCD-induced ones is most pronounced. (We will defer the study of the other Types to future publications.)   

We are in particular interested in the couplings of the (pseudo)scalars to gauge bosons and the triple-Higgs couplings. The (pseudo)scalar-gauge couplings 
$\lambda_{HAZ}$ and $\lambda_{HH^+W^-}$ are proportional to $\sin(\beta-\alpha)$, and $\lambda_{hAZ}$ and $\lambda_{hH^+W^-}$ to $\cos(\beta-\alpha)$, while $\lambda_{AH^+W^-}$ is independent of the 2HDM angles. 
The LHC data requires at least one of $h$ and $H$ to have a mass near 125\,GeV and SM-like couplings. In order for $h$ to satisfy this condition, $|\sin(\beta-\alpha)|~(|\cos(\beta-\alpha)|)$ should not be too far from 1 (0). This implies that couplings proportional to $\sin(\beta-\alpha)$ should be larger than those proportional to $\cos(\beta-\alpha)$, which indeed vanishes in the decoupling limit \cite{Gunion:2002zf}.\footnote{The decoupling limit, $\cos(\beta-\alpha)\to 0$, means that $h$ has a mass near 125 GeV and very SM-like coupling strengths, while all the other states are much heavier.} However,  given the current measurements of the properties of the $\hobs$, this limit need not be strictly adhered. For this reason, we treat $\sin(\beta-\alpha)$ as a free parameter here.

As for the triple-Higgs couplings, the CP-conserving model we are considering here contains eight of these, namely $\lambda_{hhh}$, $\lambda_{hhH}$, $\lambda_{hHH}$, $\lambda_{HHH}$, $\lambda_{hAA}$, $\lambda_{HAA}$, $\lambda_{hH^+H^-}$ and $\lambda_{HH^+H^-}$. The explicit expressions for these couplings are more complicated than for the (pseudo)scalar-gauge ones above. They are all functions of both $\sin(\beta-\alpha)$ and $\cos(\beta-\alpha)$, as well as of the quartic $\lambda_i$ parameters from the scalar potential in Eq.~(\ref{eq:2hdmpot}). However, all the $\lambda_i$ dependence can be written in terms of their combinations that are invariant under $U(2)$ basis changes in the potential. Thus the only basis dependence of these couplings comes from the angles. For explicit expressions, see Ref.~\cite{Gunion:2002zf}. 

\section{Parameter space scans and constraints}\label{sec:scans}

\begin{table}[t!]
\centering\begin{tabular}{|c|c|}
\hline
Observable & Measurement  \\
\hline \hline
${\rm BR}(B\to X_s \gamma) \times 10^{4} $   & $3.32\pm0.15$~\cite{Amhis:2016xyh}	\\ 
${\rm BR}(B_u\to \tau^\pm \nu_\tau) \times 10^{4} $ & $1.06\pm0.19$~\cite{Amhis:2016xyh}	\\ 
${\rm BR}(B_s \to \mu^+ \mu^-)\times 10^{9} $ & $3.0\pm 0.85$~\cite{Aaij:2017vad}\\ 
\hline
$\mu_{\gamma\gamma}$ & $1.14^{+0.19}_{-0.18}$~\cite{Khachatryan:2016vau}\\
$\mu_{ZZ}$ & $1.29^{+0.26}_{-0.23}$~\cite{Khachatryan:2016vau}\\
$\mu_{WW}$ & $1.09^{+0.18}_{-0.16}$~\cite{Khachatryan:2016vau}\\
$\mu_{\tau\tau}$  & $1.11^{+0.24}_{-0.22}$~\cite{Khachatryan:2016vau} \\
$\mu_{bb}$ & $0.70^{+0.29}_{-0.27}$~\cite{Khachatryan:2016vau} \\
\hline 
\end{tabular}
\caption{\label{tab:Higgs-rates} Measured values of the $B$-physics observables and $\hobs$ signal rates imposed as constraints on the scanned points.}
\end{table}

We numerically scanned the parameters of the Type-I 2HDM using the 2HDM Calculator (2HDMC)~\cite{Eriksson:2009ws} in the ranges:
\begin{center}
		$m_H$: 150 -- 750\,GeV\,;~~$m_{\hpm}$: 50 -- 750\,GeV\,;~~$m_A$: 50 -- 750\,GeV\,;\\
		$\sin(\beta-\alpha)$: $-1$ -- 1\,;~~$m_{12}^2$: 0 -- $m_{\A}^2\sin\beta\cos\beta$\,;~~$\tanb$: 2 -- 25\,, \\
\end{center}
with $m_h$ fixed to 125\,GeV and $\lambda_6\,,\lambda_7$ to zero, such that each point satisfied the following set of requirements. 

\begin{itemize}
\item Unitarity (default unitarity limit is $16\pi$), perturbativity (default perturbativity limit is $4\pi$) and Higgs potential stability conditions were enforced with methods provided by  2HDMC.

\item The oblique parameters $S$, $T$ and $U$ were calculated with 2HDMC methods and were required to fall within the 95\% Confidence Level (CL) ellipsoid based on 2018 PDG values~\cite{Tanabashi:2018oca}:
\bea
S &=& 0.02 \pm 0.10\,,\\
T &=& 0.07 \pm 0.12\,,\\
U &=& 0.00 \pm 0.09\,,
\eea
with correlations $\rho_{ST}=0.92$, $\rho_{SU}=-0.66$ and $\rho_{TU}=-0.86$.

\item All scalar states in the models satisfied all ($95\%$ CL) constraints included in the program HiggsBounds 5.2.0 \cite{Bechtle:2013wla}.

\item The $B$-physics observables were calculated with SuperIso 3.4 \cite{Mahmoudi:2008tp}. They were required to meet the limits from the SuperIso manual (95\% CL), except for the three Branching Ratios (BRs) listed in  Tab.~\ref{tab:Higgs-rates}, for which we applied the constraints on the $m_{H^\pm},\,\tan\beta$ plane derived in~\cite{Mahmoudi:2017mtv}. 

\item The signal strengths for $h\to\gamma\gamma$, $ZZ$, $WW$, $\tau\tau$ and $b\bar{b}$, calculated using HiggsSignals 2.2.0 \cite{Bechtle:2013xfa}, were required to lie within $2\sigma$ of the LHC measurements for $\hobs$ given in Tab.~\ref{tab:Higgs-rates}.

\end{itemize}

We point out here that due to the absence of a dark matter (DM) candidate particle, the constraints from the relic abundance of DM and from the experimental facilities for its detection are irrelevant in the 2HDM. Such constraints would indeed apply in the case of the Minimal Supersymmetric Standard Model, which contains two Higgs doublets as well, and also predicts fermionic DM. The prospects of the pair-production of the heavy Higgs bosons in this model, with one of these decaying into the DM itself, have been studied in, e.g., \cite{Arganda:2017wjh}. 

\section{Results and discussion}
\label{sec:results}

For each scanned point, we  calculated tree-level cross sections in $pp$ collisions with $\sqrt{s}=13$\,TeV for all possible $q\bar q^{(\prime)}\to h_i h_j$ processes, with $h_{i,j} = (h,H,A,H^\pm)$. These cross sections were calculated using the 2HDMC model \cite{Eriksson:2009ws} with MadGraph5\_aMC@NLO \cite{Alwall:2014hca}. For the neutral 2-body final states (2BFSs), we also calculated the cross sections for $b\bar{b}\to h_i h_j$ in the five flavour scheme using the same methods and for $gg\to h_i h_j$ (gluon-gluon fusion) using MadGraph based codes \cite{Hespel:2014sla}. 

From these, we estimated cross sections for the 3-Body Final States (3BFSs) $h_i + h_j + h_k/V_k$, with $h_i = (h, H, A, H^\pm)$ and $V = (W^\pm, Z)$. This was done by multiplying the cross section for a given $2\to 2$ process (where available) with the appropriate BR, considering all possible on-shell decays of the heavier (pseudo)scalars. 
We note that the majority of points accepted in our scan contain states whose widths are several orders of magnitude smaller than their masses, so we do not expect large corrections due to our narrow-width approximation.\footnote{Our scan does contain a minority of points for which $A$ and/or $H^\pm$ have large widths. However, the large cross sections highlighted in the following sections all correspond to decays of states with narrow widths.}
Furthermore, while a full analysis would take into account all the contributions, including the interference effects among different channels, to the production of a given 3BFS simultaneously, we consider the contribution of each channel separately here. We are afforded this simplification by the fact that the 3BFS cross sections presented in the following sections are typically dominated by a single process.

\subsection{Charged final states}

\begin{figure}[t!]
\includegraphics[angle=0,width=0.33\textwidth]{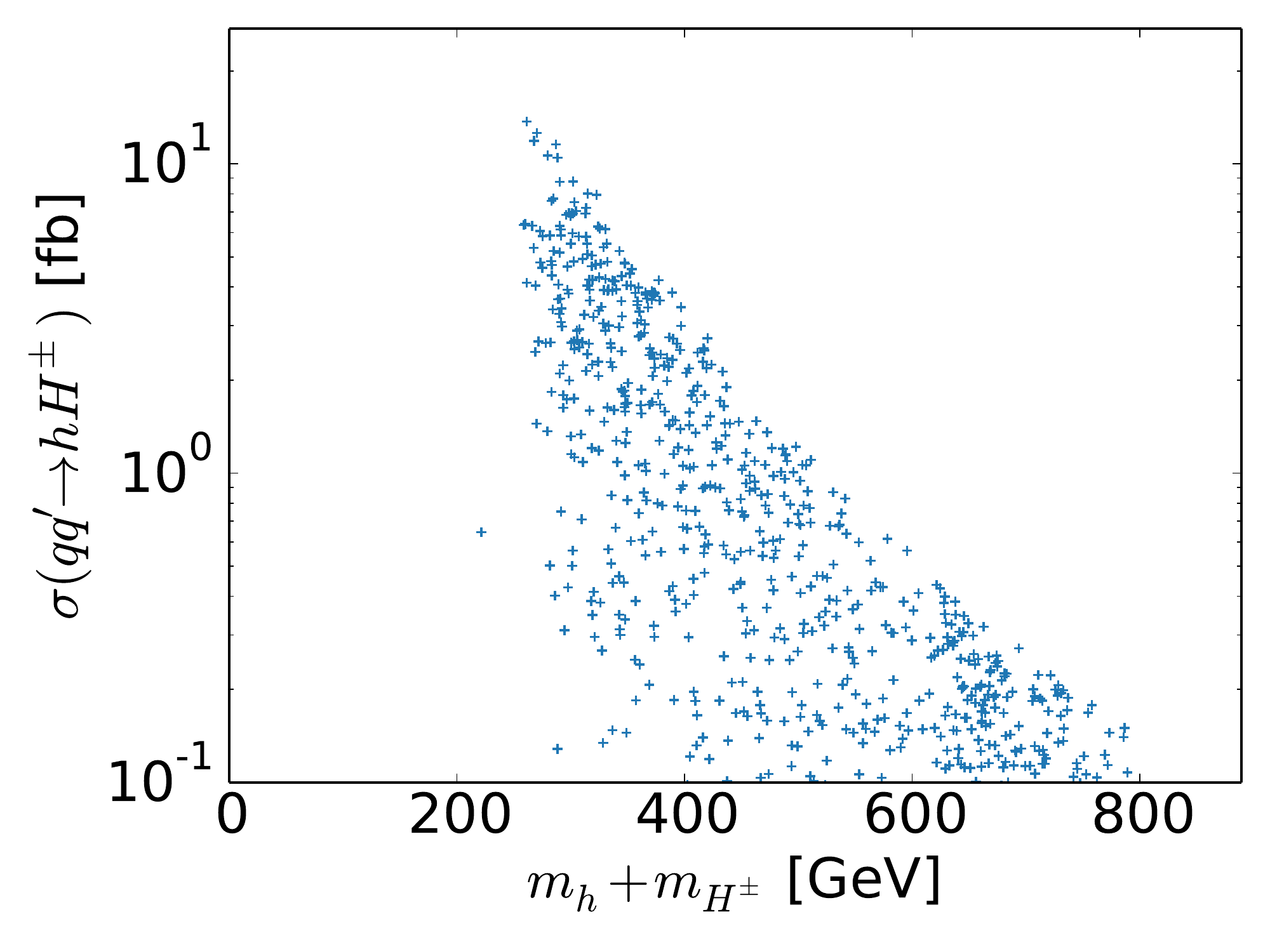}
\includegraphics[angle=0,width=0.33\textwidth]{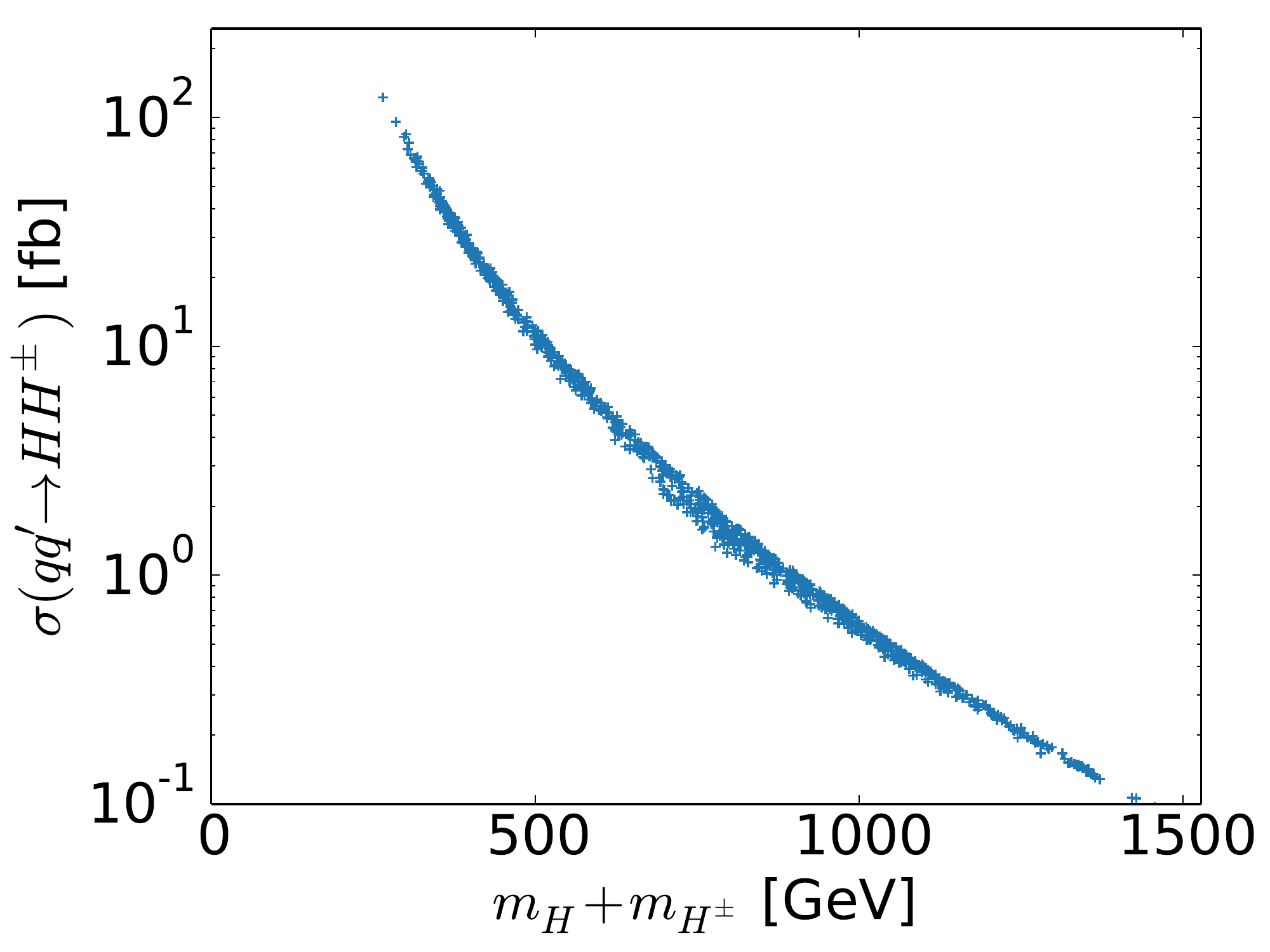}
\includegraphics[angle=0,width=0.33\textwidth]{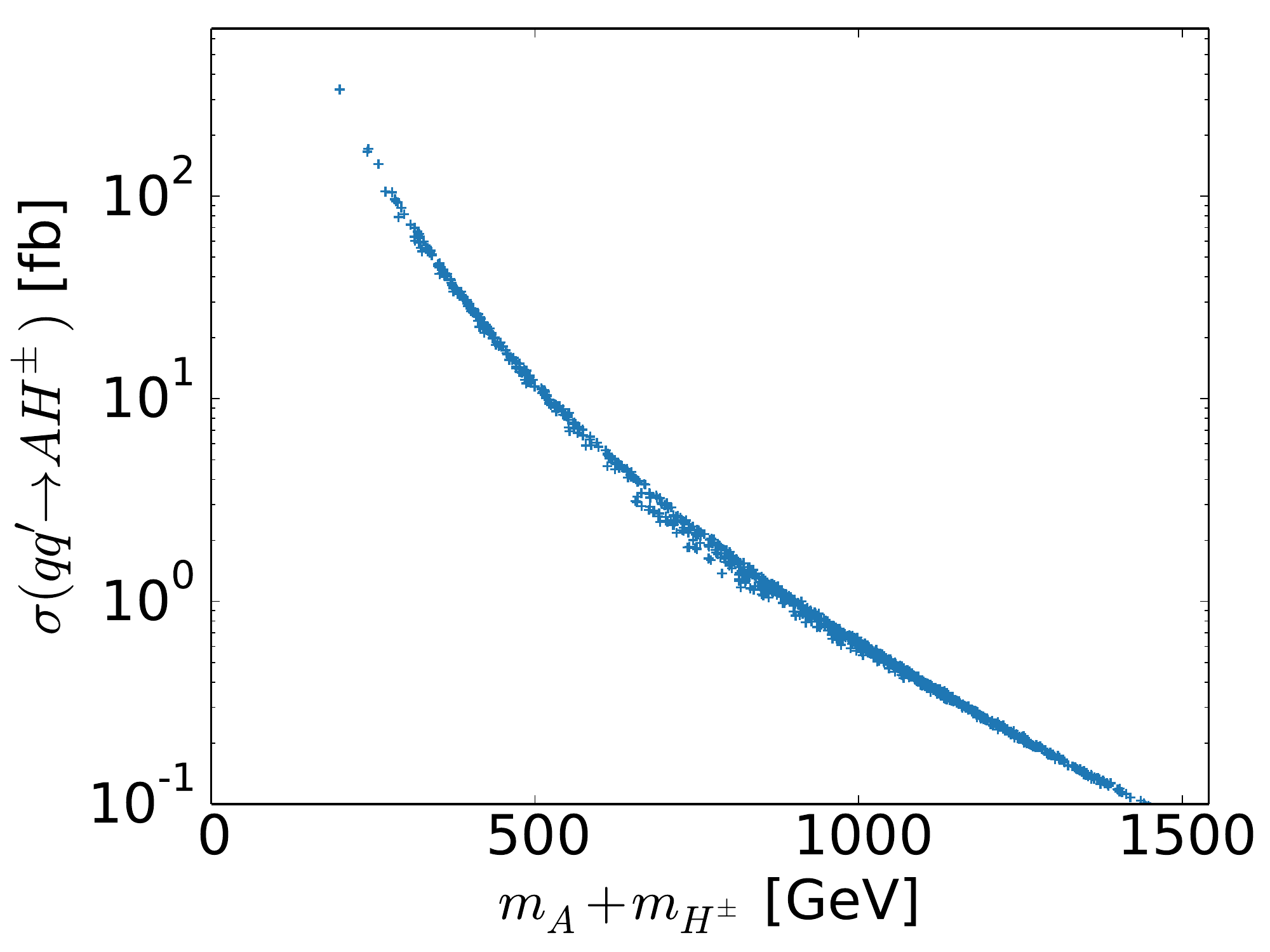}
\caption{Cross sections for the three possible charged 2BFSs.}
\label{fig:charged2bfs}
\end{figure}

The charged 2BFSs, each containing the $H^\pm$ along with one neutral Higgs state, are shown in Fig.~\ref{fig:charged2bfs}. These are all necessarily produced by an initial $q\bar{q}'$ state, having no counterpart in $gg/b\bar{b}$ production, and each shows a maximum cross section of at least 10\,fb in some kinematic regions. Whereas the cross sections for $HH^\pm$ and $AH^\pm$ states are strongly correlated with their cumulative masses, those of $hH^\pm$ show greater variation. We find that this variation is correlated with $\sin(\beta-\alpha)$, with maximal cross sections corresponding to minimal $\sin(\beta-\alpha)$. This is consistent with a cross section dominated by an $s$-channel $W^\pm$, whose coupling to $hH^\pm$ is proportional to $\cos(\beta-\alpha)$, as noted earlier. Because the $h$ is required to have very SM-like properties, the points selected by our scans have $\left|\sin(\beta-\alpha) \right |$ close to 1, which means that $\cos(\beta-\alpha)$ may span several orders of magnitude, resulting in large variation in possible $hH^\pm$ cross sections.  Conversely, the $\lambda_{HH^+ W^-}$ coupling varies as $\sin(\beta-\alpha)$ and the $\lambda_{AH^+ W^-}$ coupling has no dependence on $\sin(\beta-\alpha)$, so the cross sections for other charged 2BFSs are also consistent with dominant $s$-channel $W^\pm$ production, being determined almost entirely by the final state kinematics.

\begin{figure}[t!]
\includegraphics[width=0.33\textwidth]{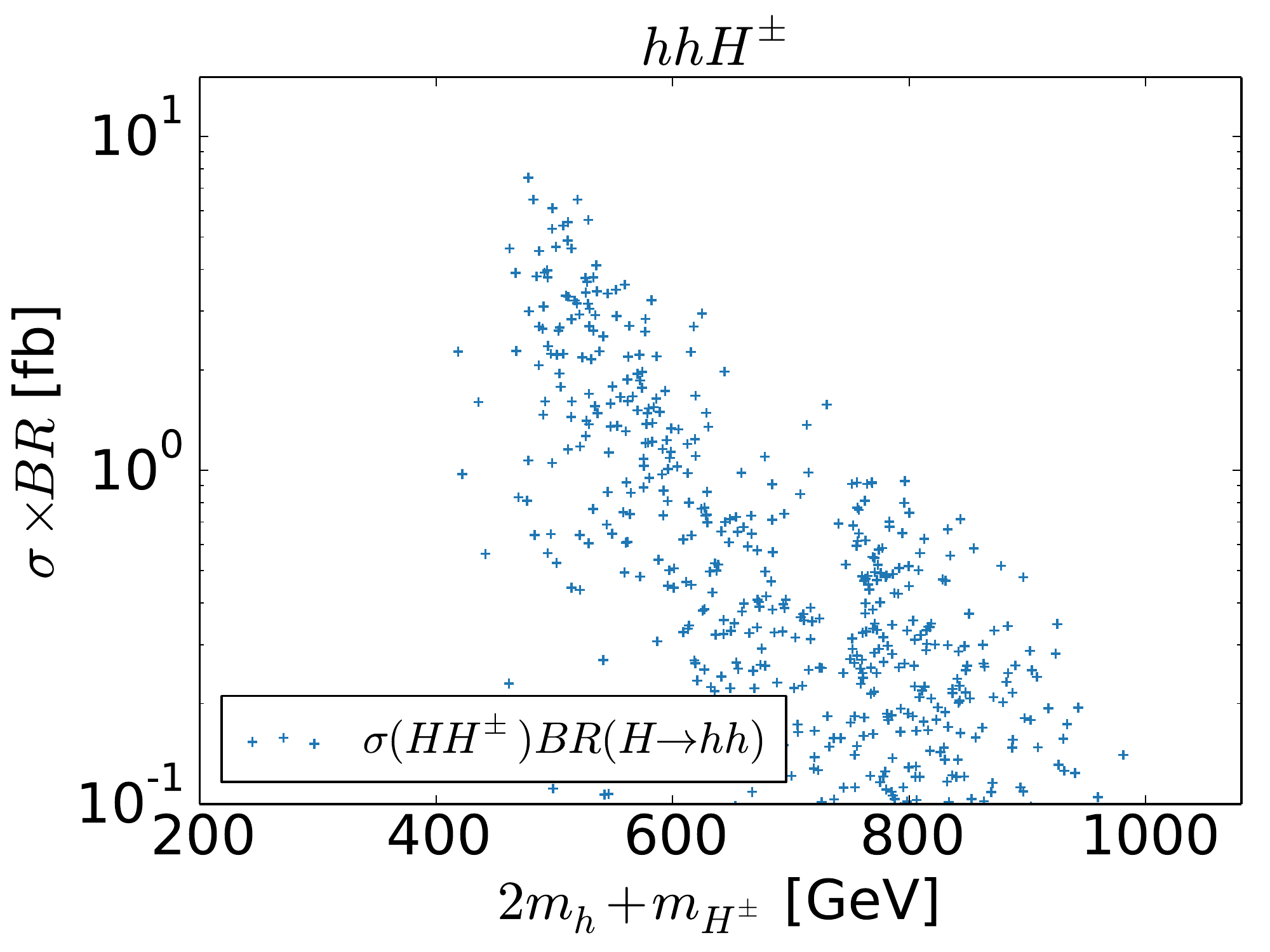}
\includegraphics[width=0.33\textwidth]{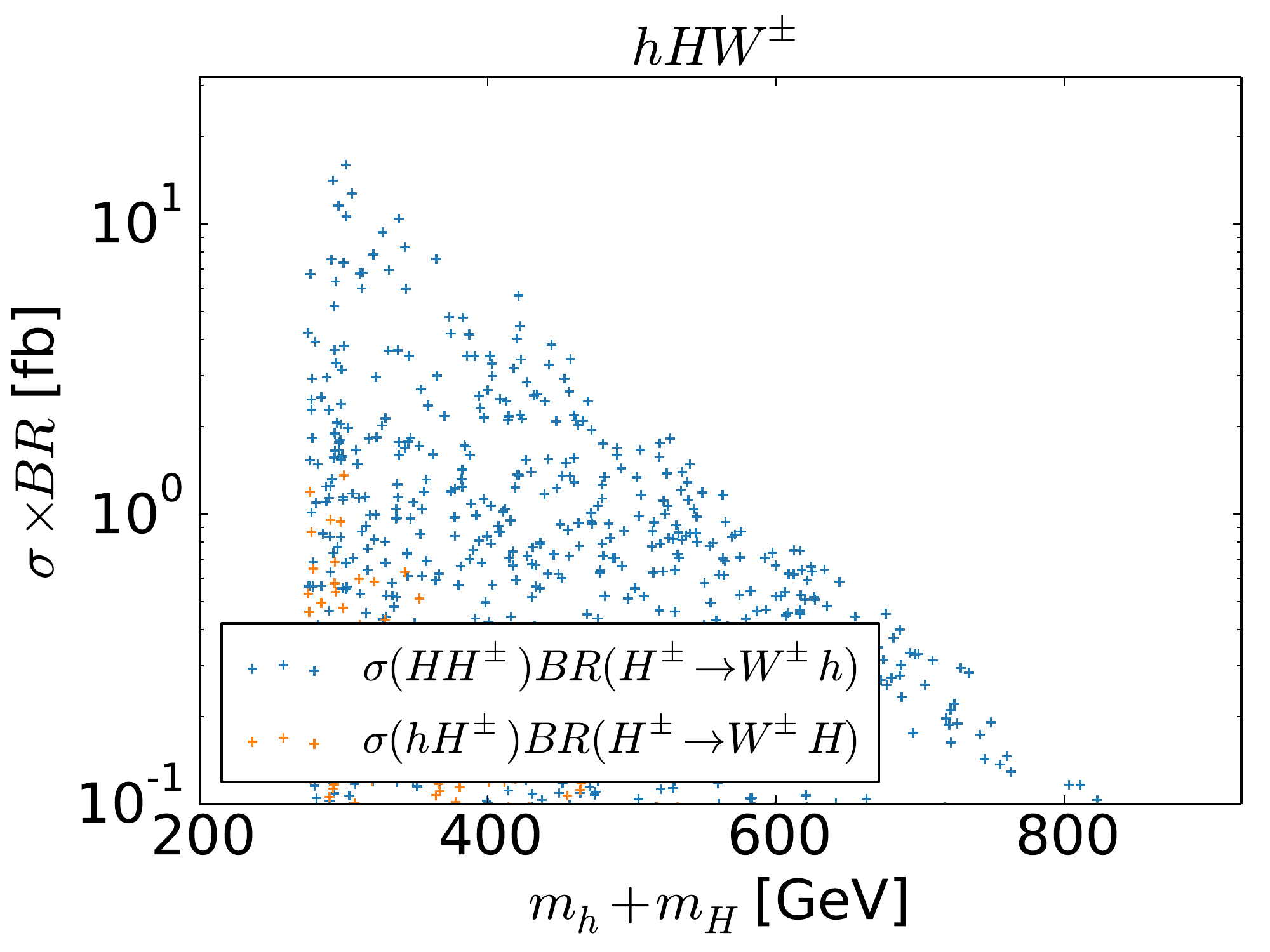}
\includegraphics[width=0.33\textwidth]{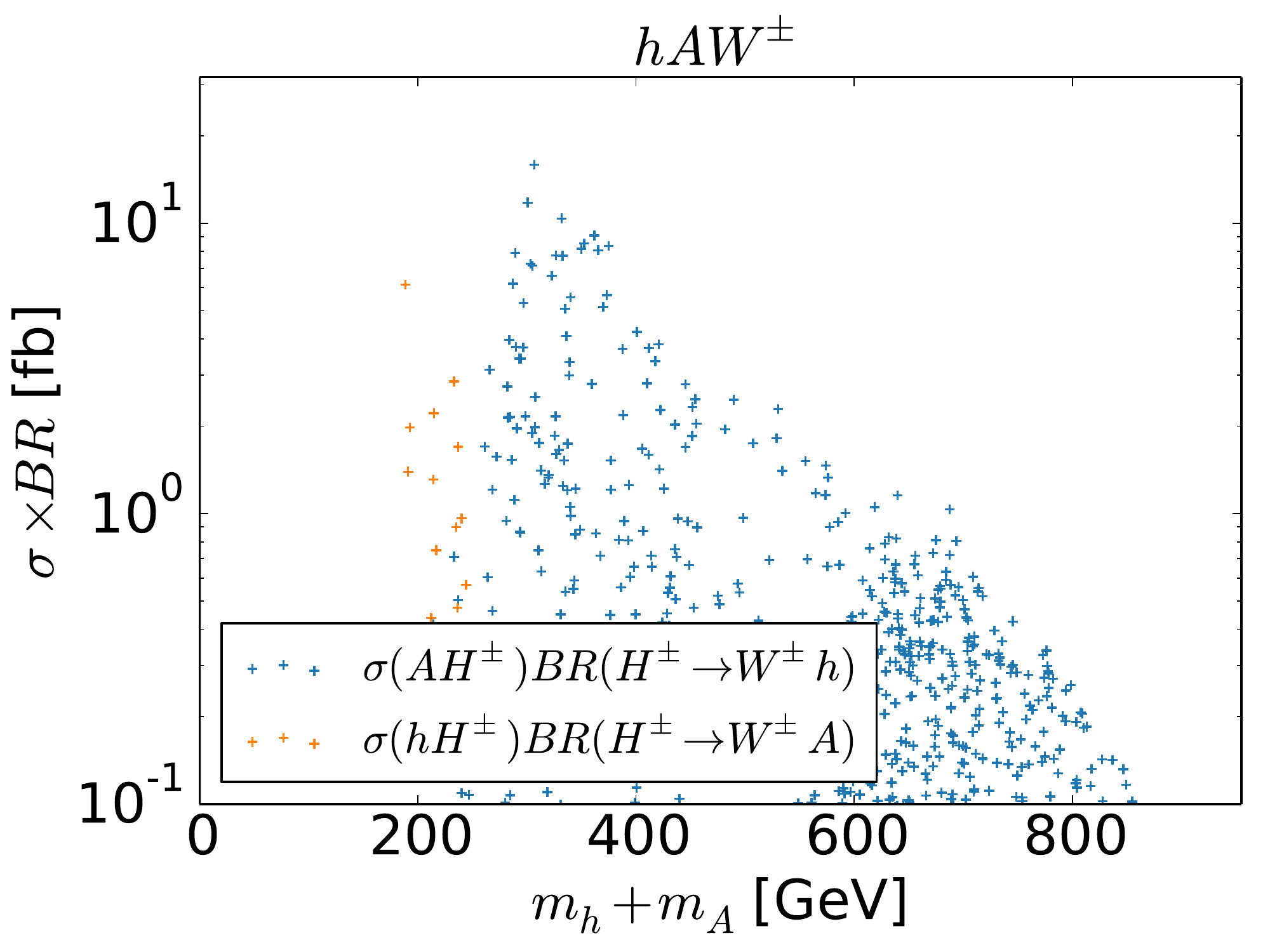}
\includegraphics[width=0.33\textwidth]{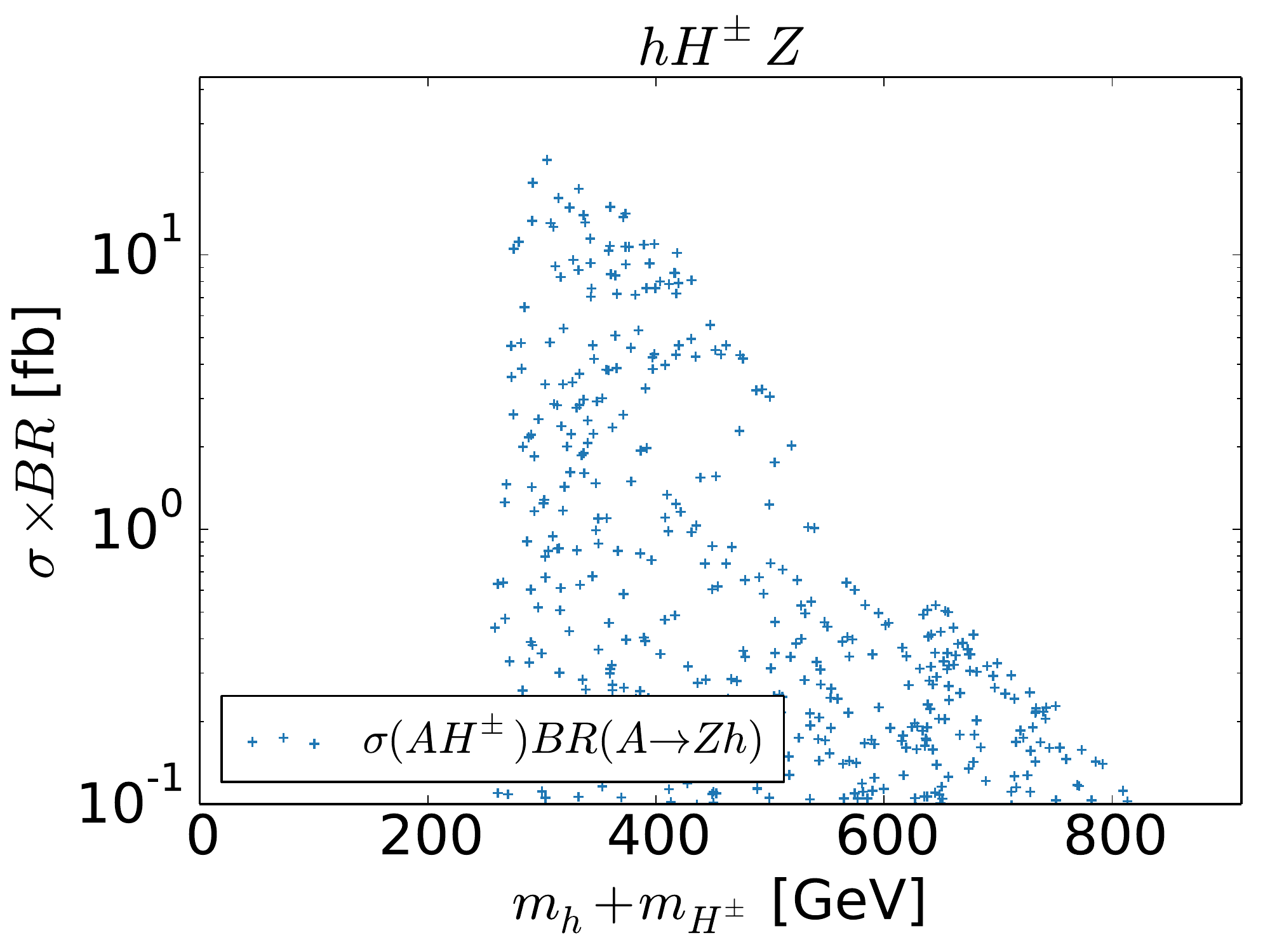}
\includegraphics[width=0.33\textwidth]{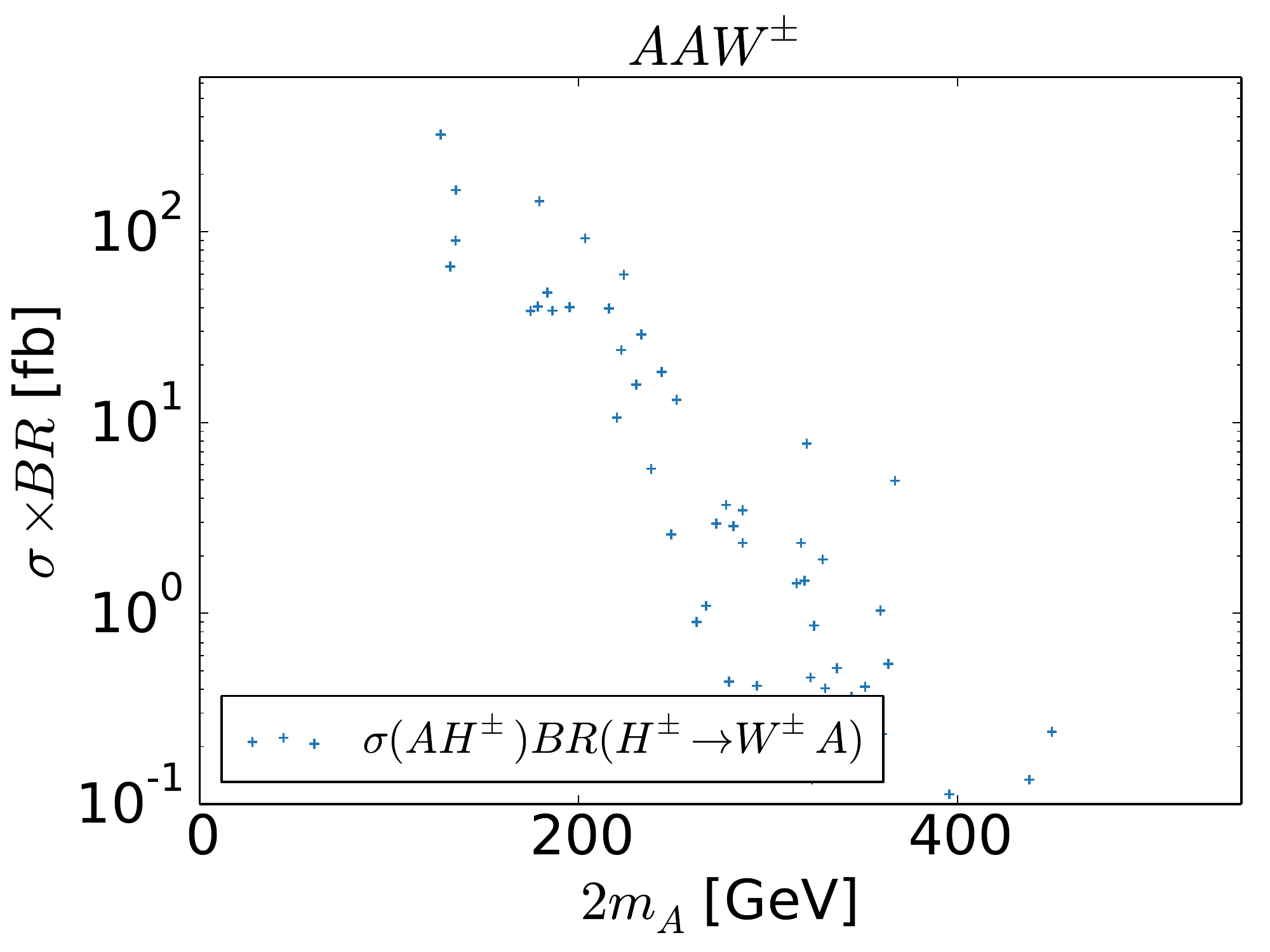}
\includegraphics[width=0.33\textwidth]{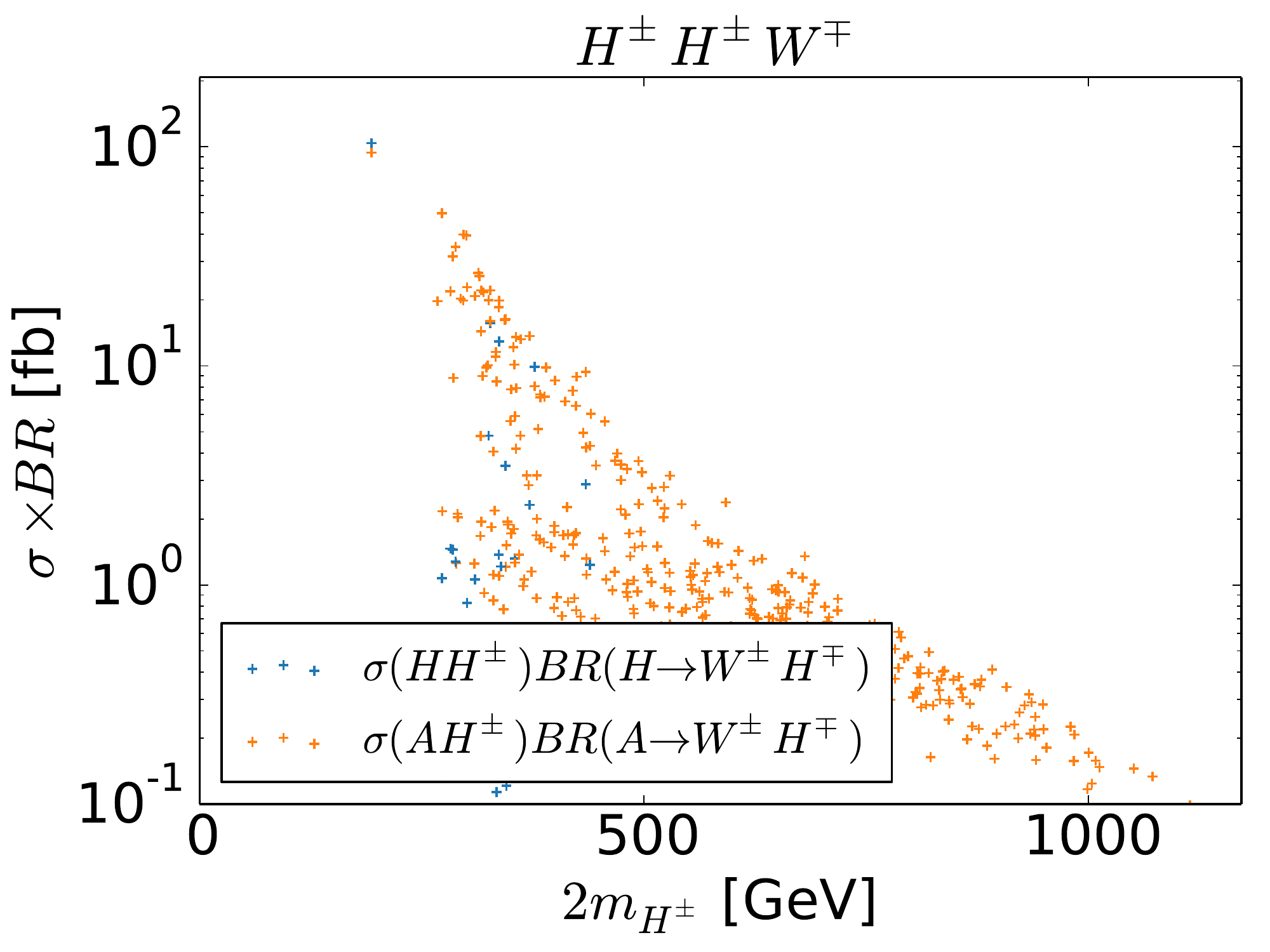}
\includegraphics[width=0.33\textwidth]{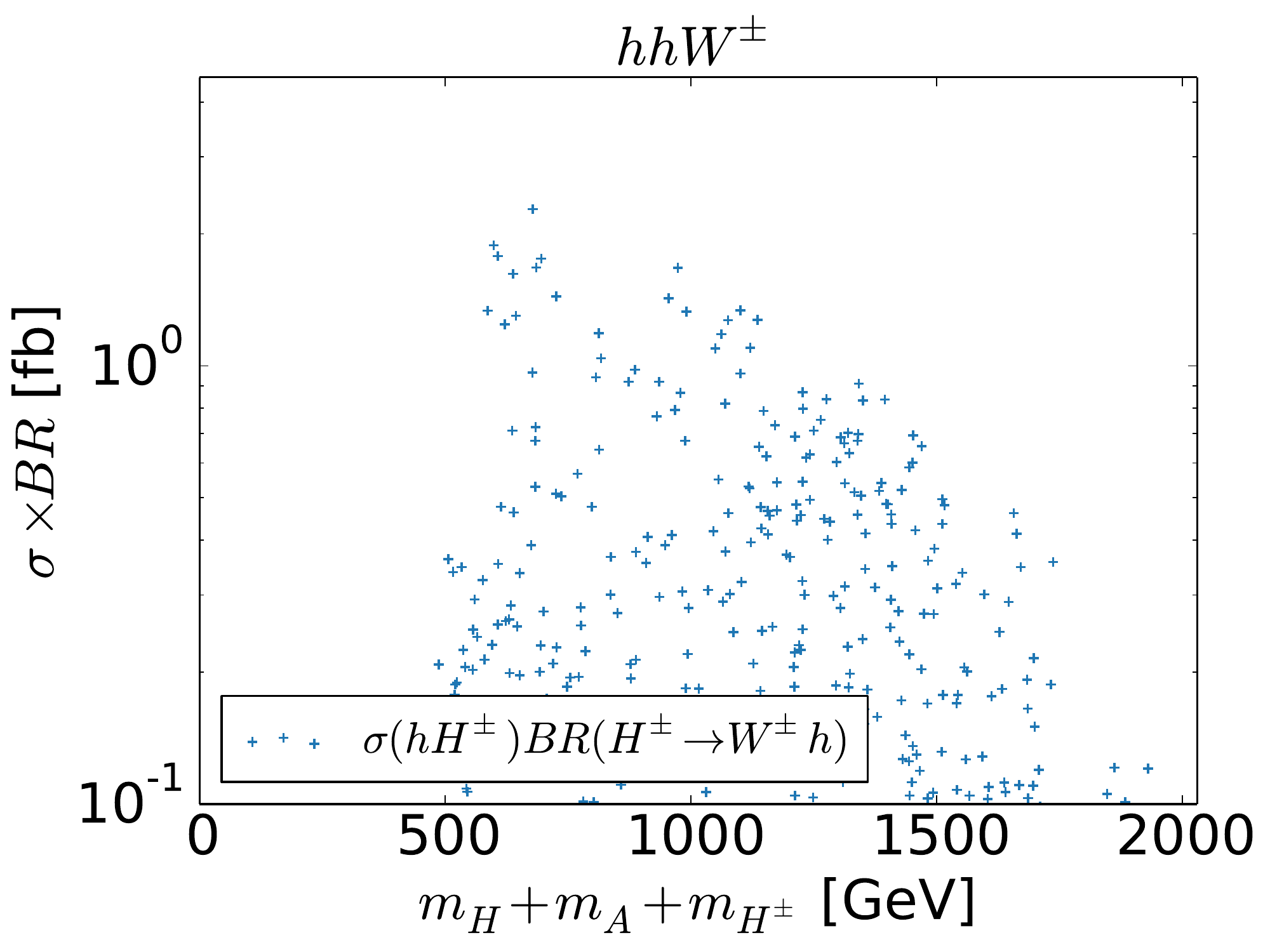}
\includegraphics[width=0.33\textwidth]{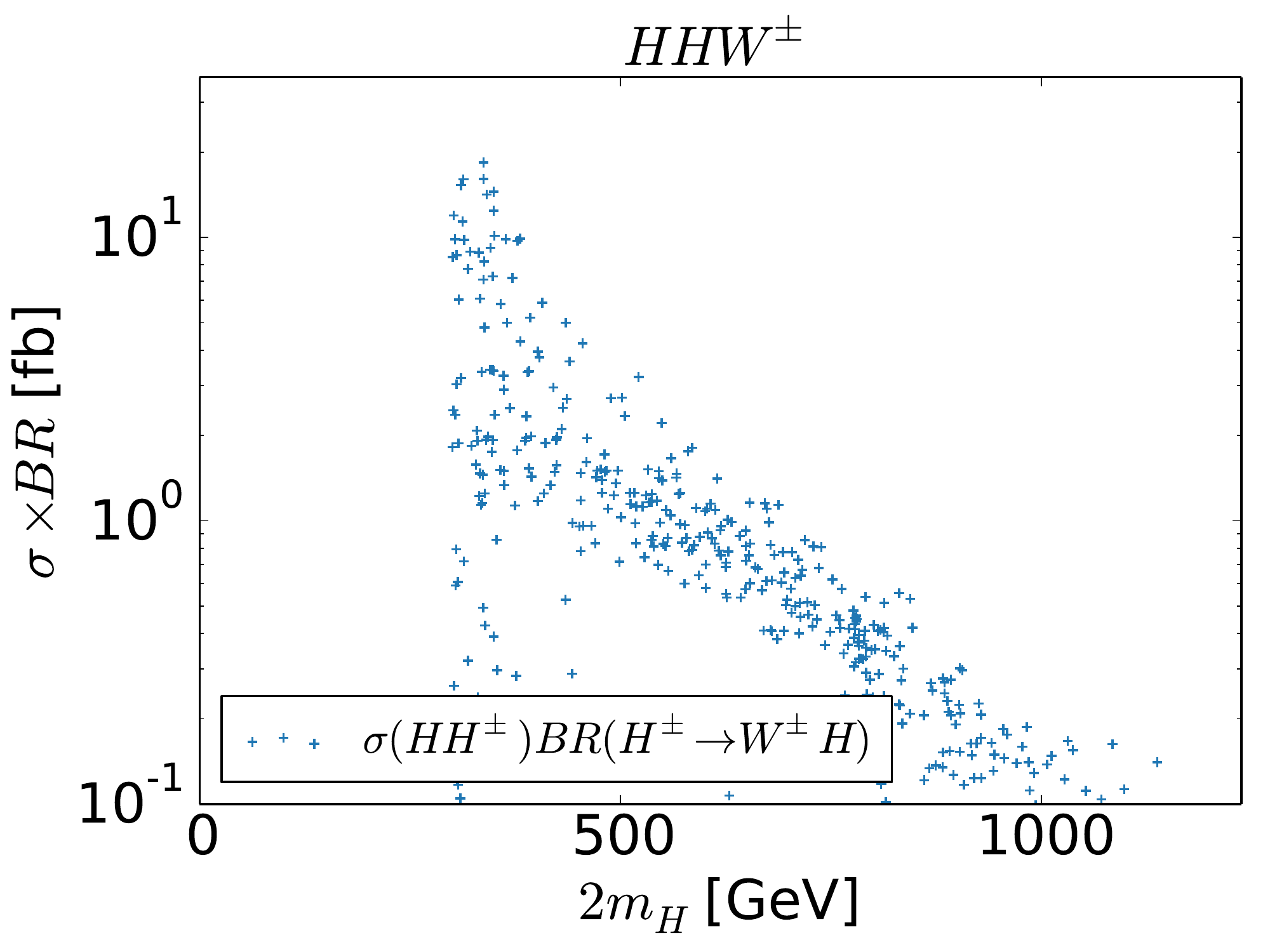}
\includegraphics[width=0.33\textwidth]{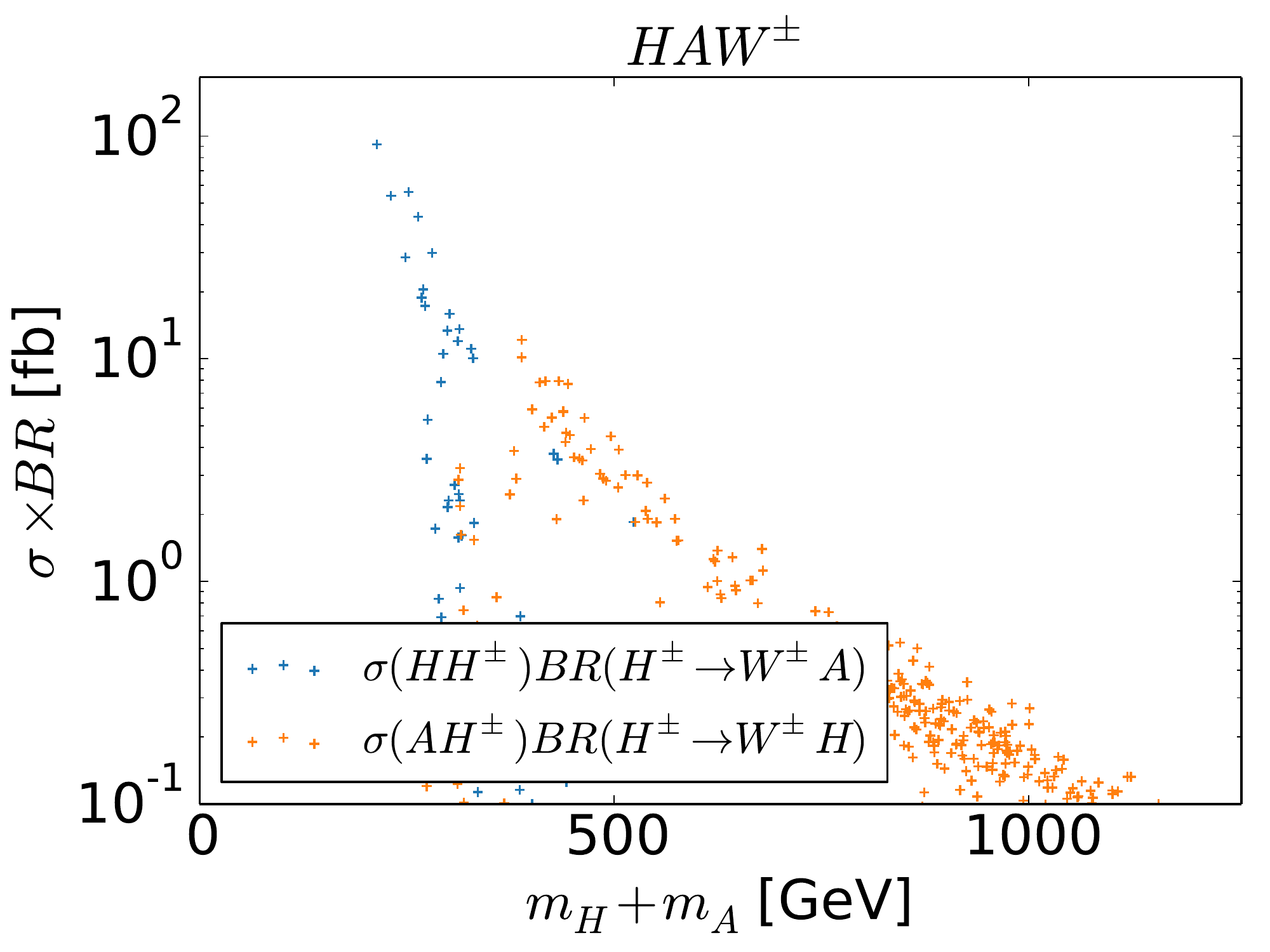}
\includegraphics[width=0.33\textwidth]{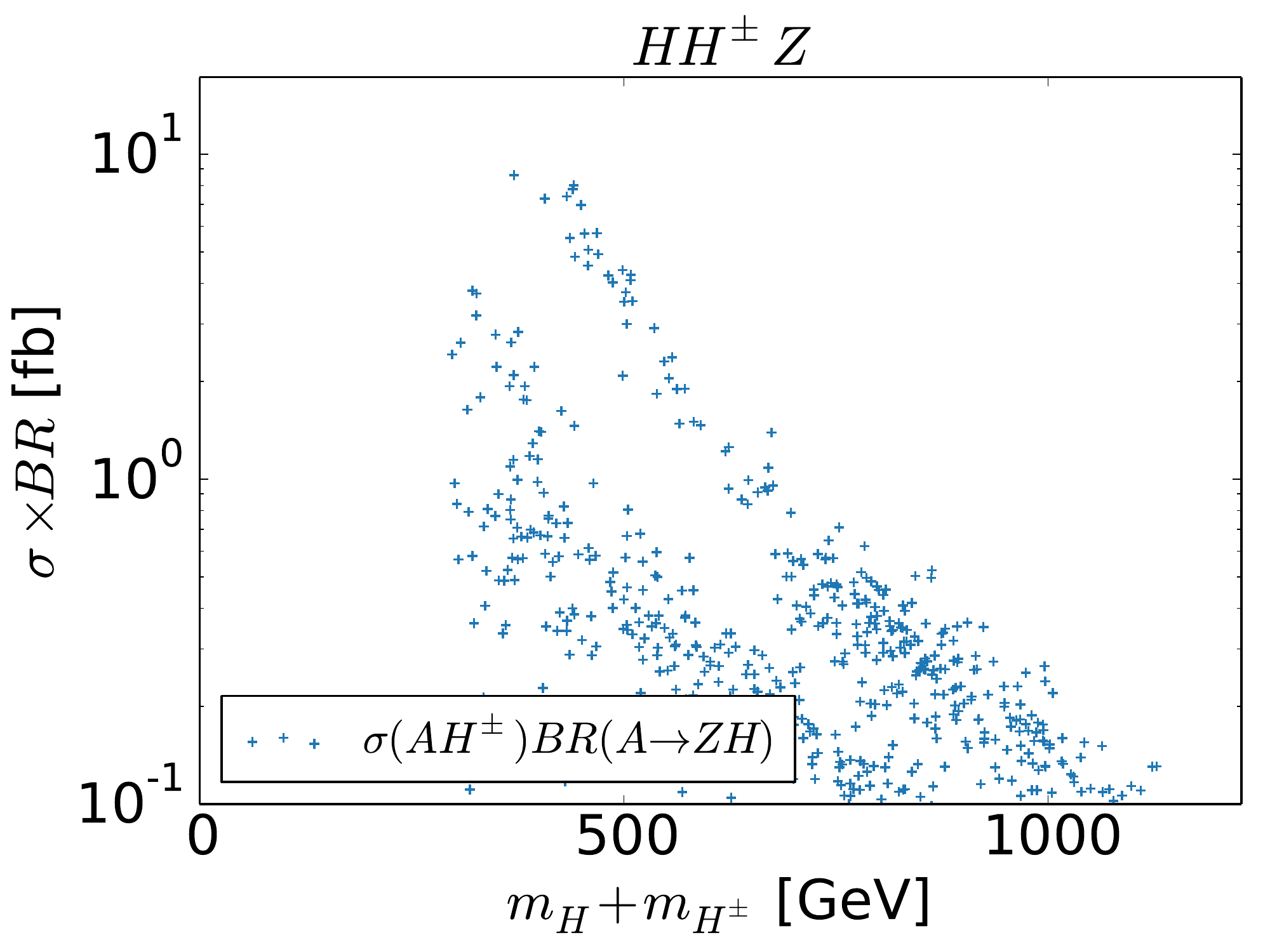}
\includegraphics[width=0.33\textwidth]{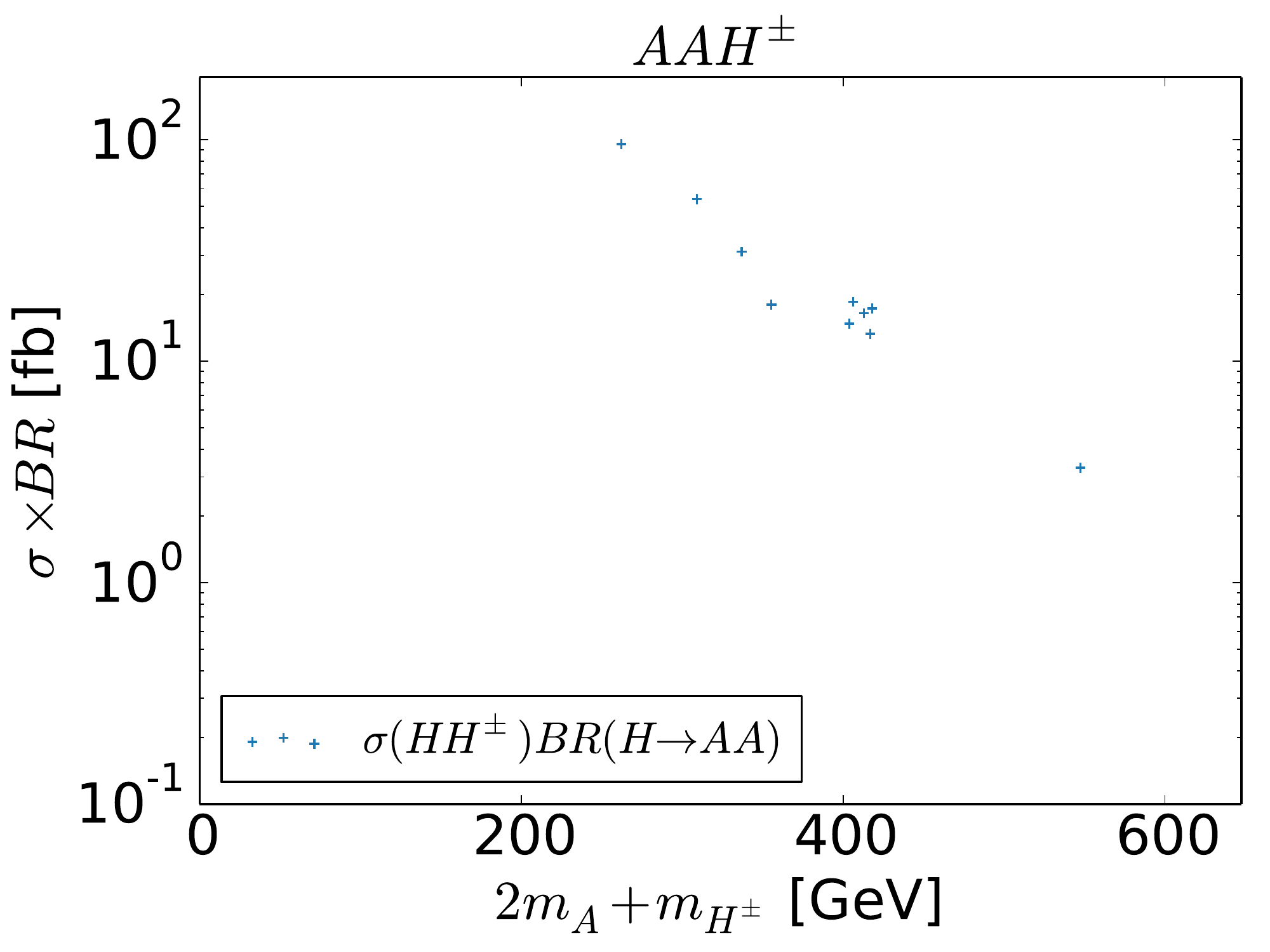}
\includegraphics[width=0.33\textwidth]{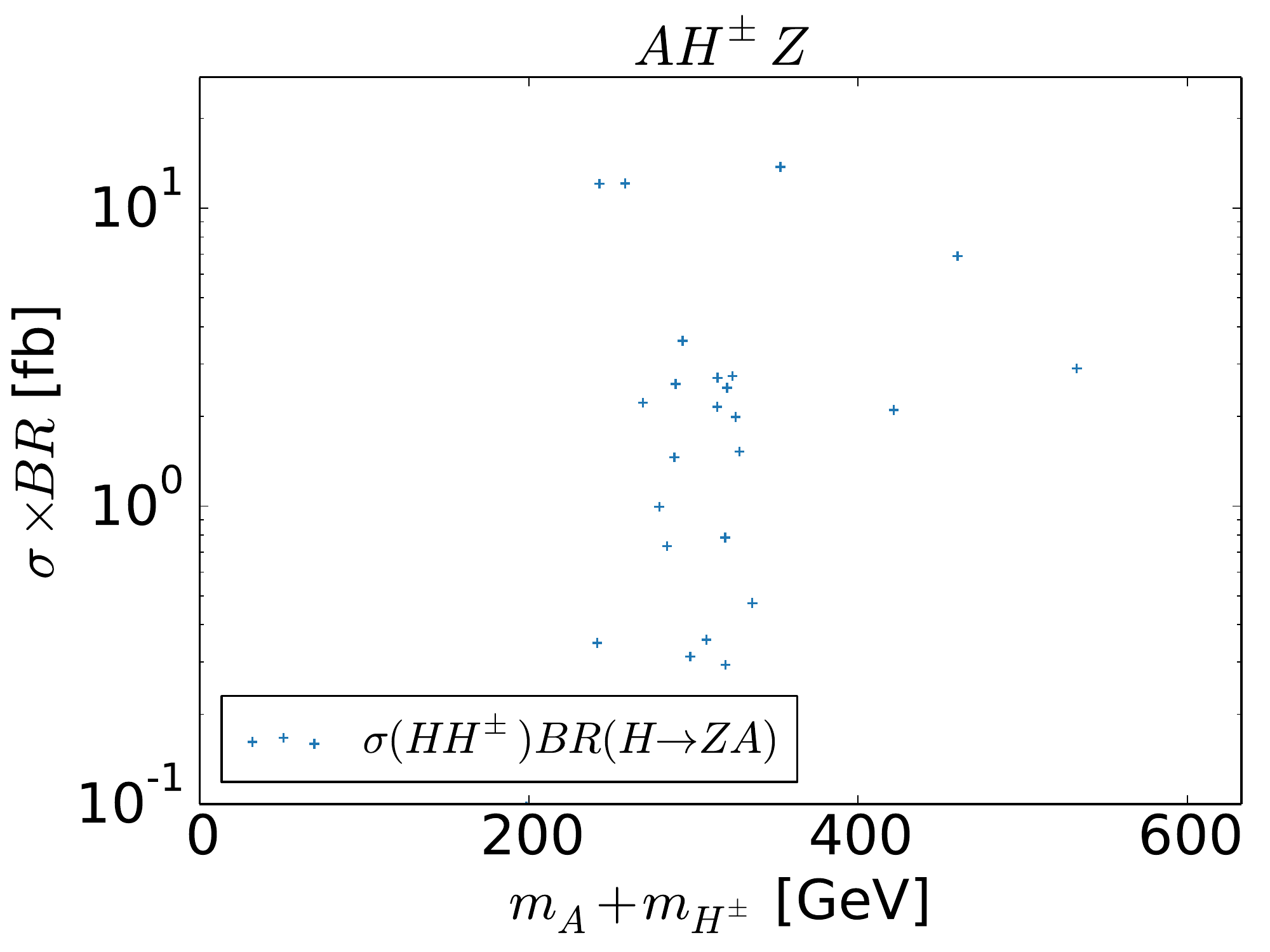}
\caption{Cross sections of $qq'$-initiated subprocesses for selected charged 3BFSs.}
\label{fig:charged3bfs}
\end{figure}

If we consider the possibility of either the charged or neutral Higgs in a 2BFS decaying, we can have final states containing either three Higgs bosons or two Higgs bosons accompanied by one gauge boson. The cross sections for such 3BFSs, for processes where it exceeds 1\,fb for at least one point from the scan, are shown in Fig.~\ref{fig:charged3bfs}. The maximal cross sections for all such 3BFSs are summarised in Tab.~\ref{tab:maxcx}. We note that there are several possible processes which would lead to cross sections of this size, and all of the possible $h_i\to h_j+h_k/V_k$ decays are represented, excepting one; $H\to H^+ H^-$ does not appear because our scan did not select any points meeting the condition $m_H>2m_{H^\pm}$ required for this decay. We also note in Fig.~\ref{fig:charged3bfs} that there are very few points selected by our scan with large cross sections involving $H^\pm\to W^\pm A$, $H\to A A$, or $H\to A Z$ decays. Again, this is simply because most points do not have masses which satisfy the kinematic requirements for these decays. However, our broad scan did find some points where the cross sections containing these decays are very substantial and a more comprehensive scanning should find additional candidates.

\begin{figure}[tb!]
\includegraphics[angle=0,width=0.33\textwidth]{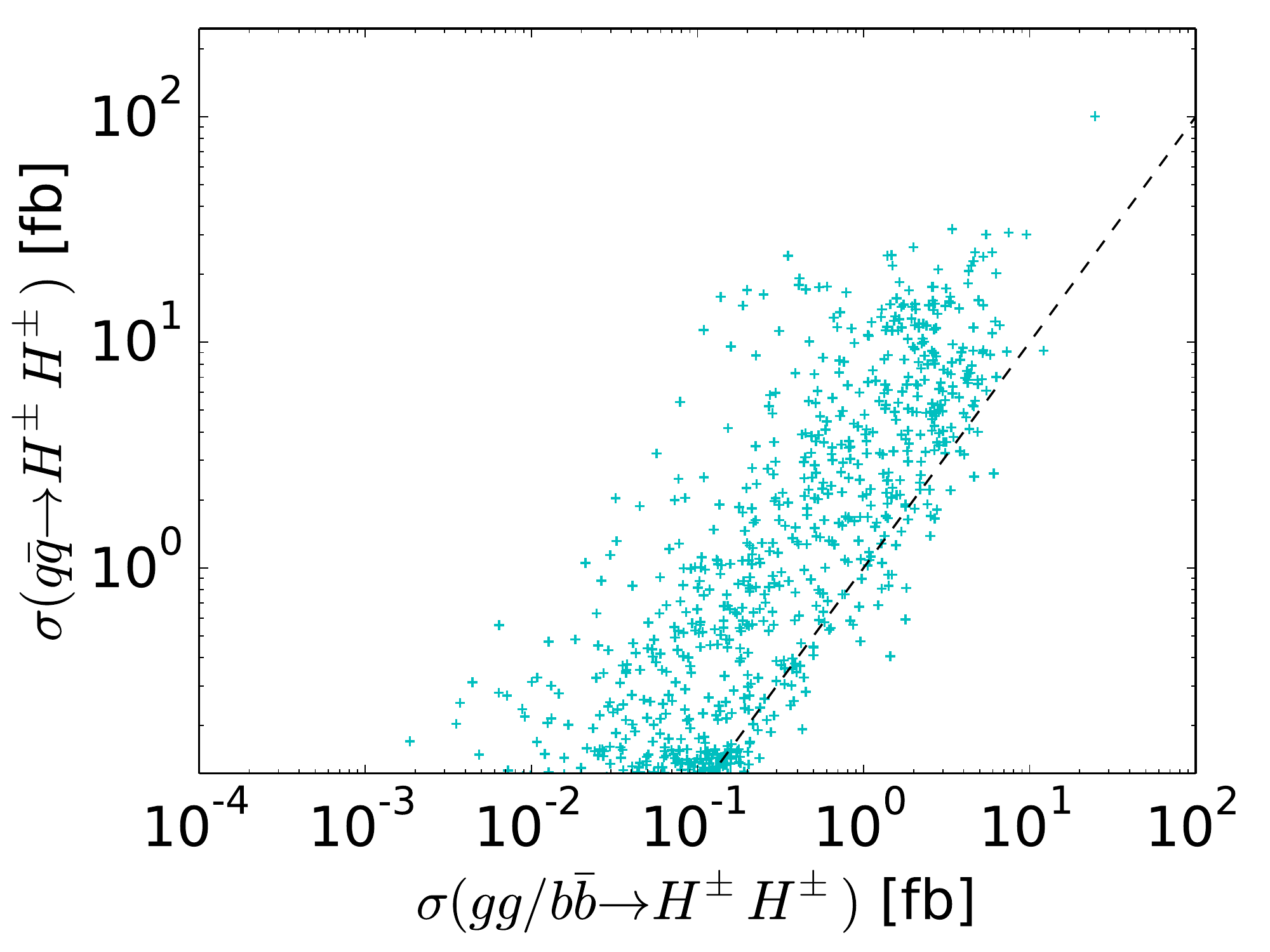}
\includegraphics[angle=0,width=0.33\textwidth]{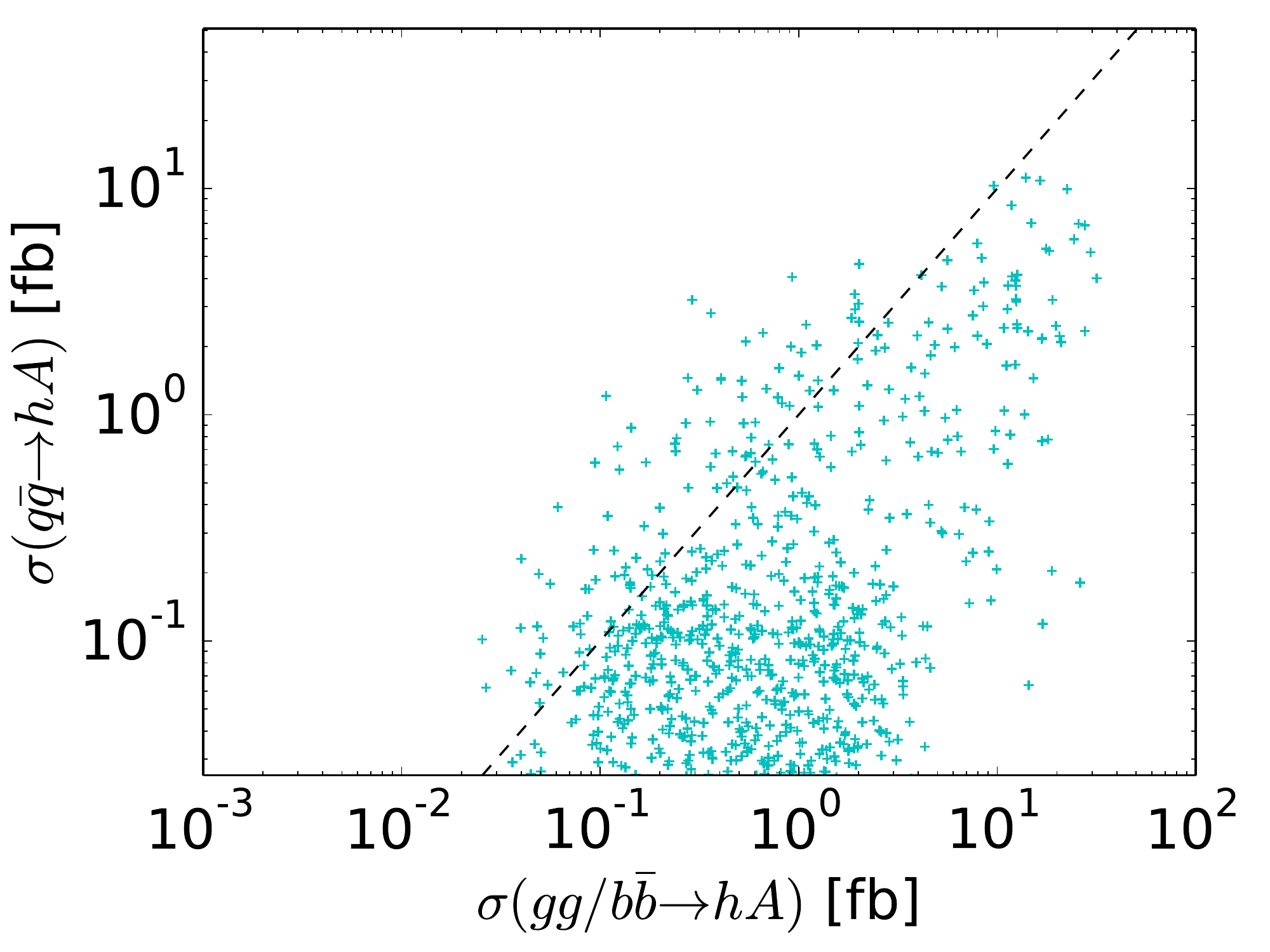}
\includegraphics[angle=0,width=0.33\textwidth]{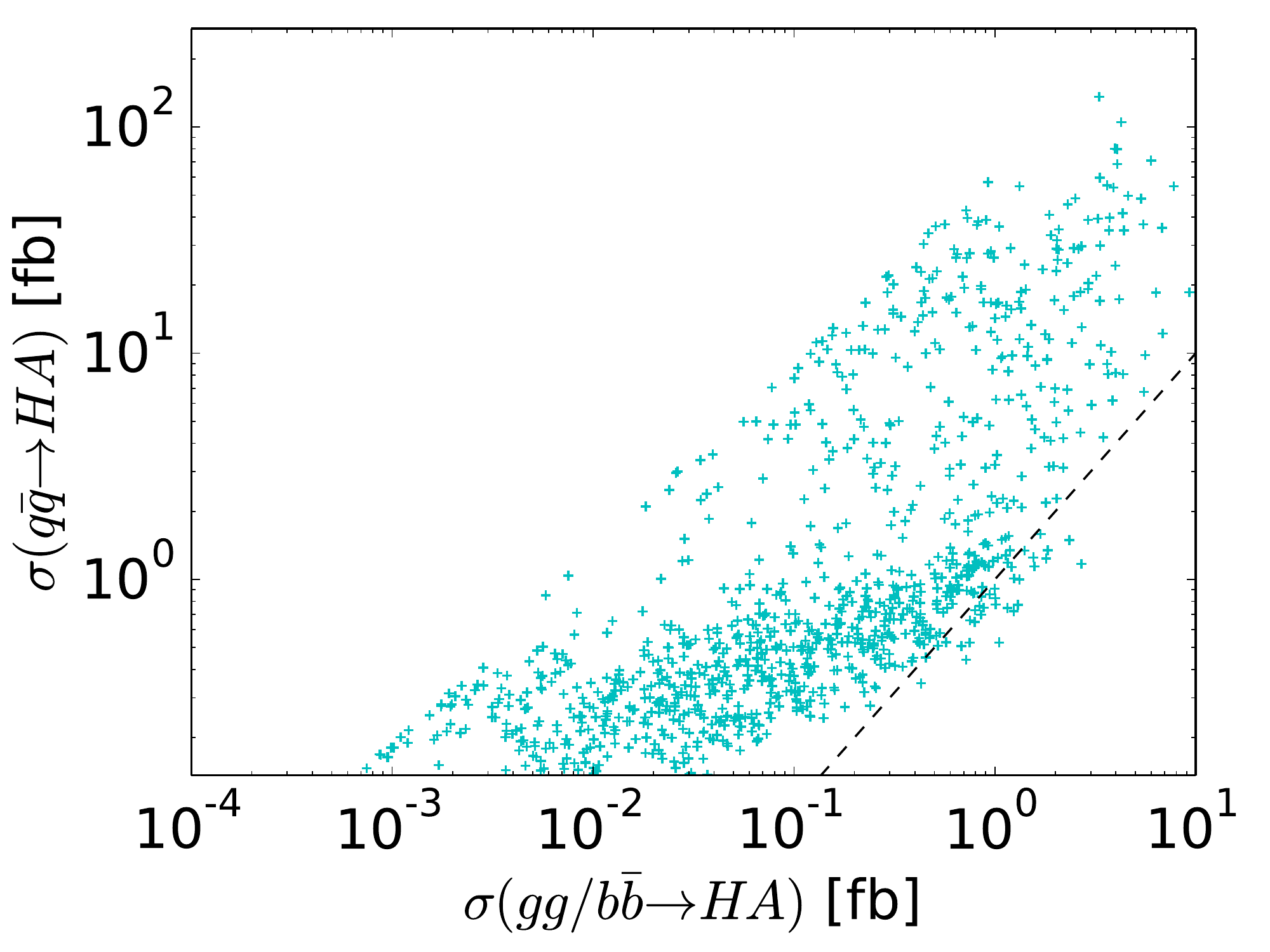}
\caption{Neutral 2BFSs for which the cross sections for $qq'$ production can exceed those for $gg/b\bar{b}$-initiated processes. The dashed line indicates where the cross sections are of equal magnitude.}
\label{fig:neutral2bfs}
\end{figure}

\begin{figure}[t!]
\begin{center}
\includegraphics[angle=0,width=0.33\textwidth]{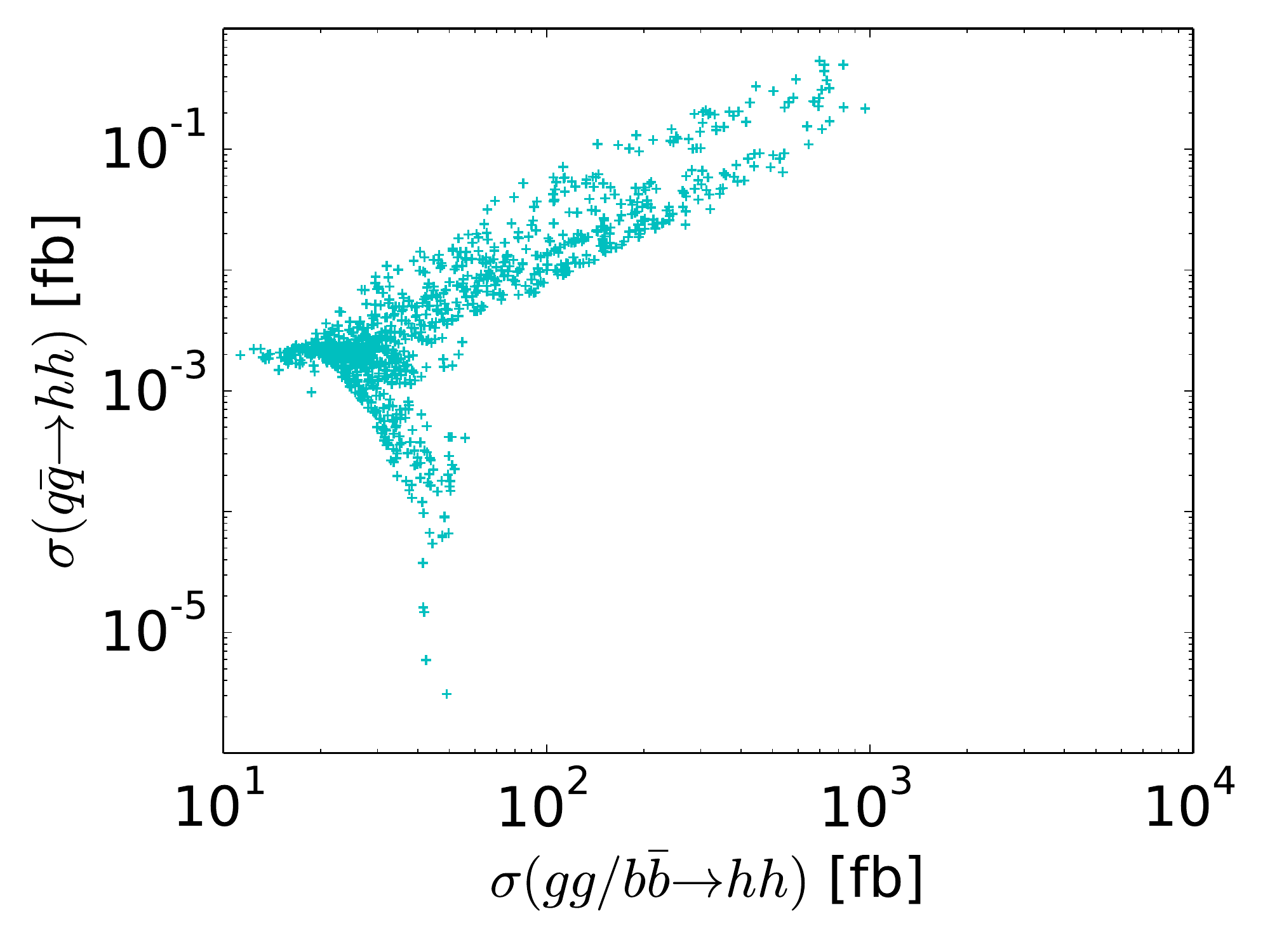}
\includegraphics[angle=0,width=0.33\textwidth]{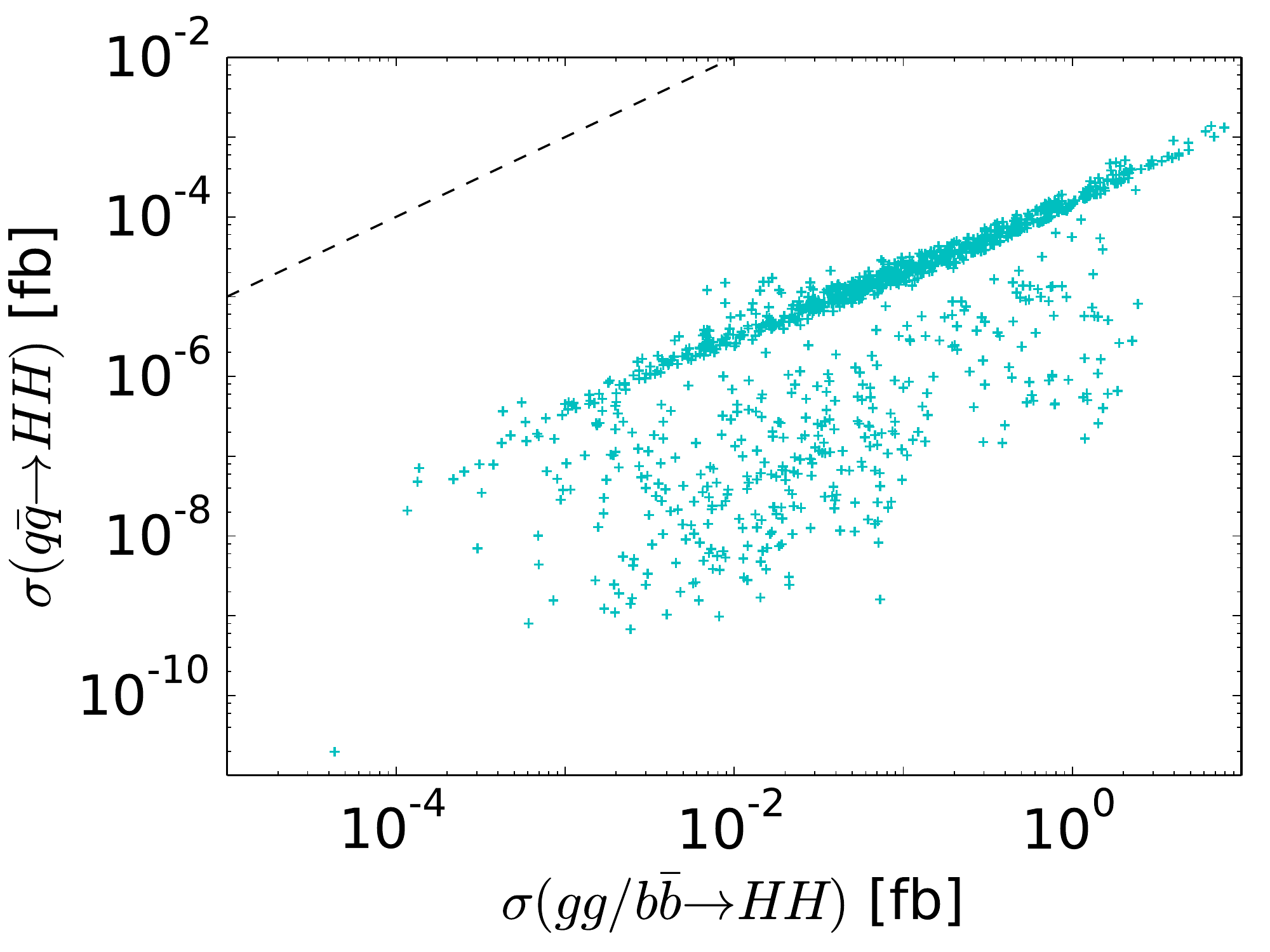}\\
\includegraphics[angle=0,width=0.33\textwidth]{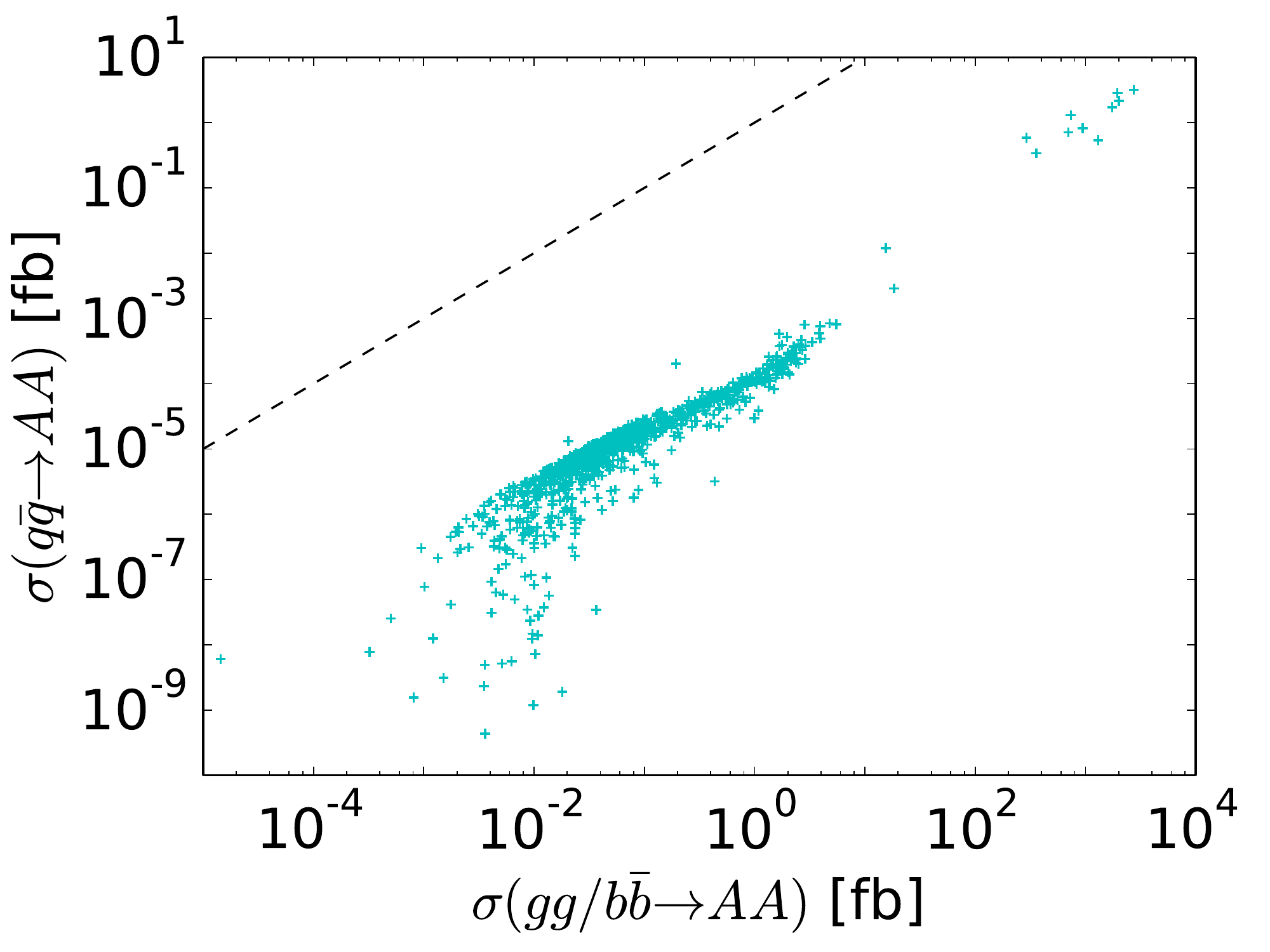}
\includegraphics[angle=0,width=0.33\textwidth]{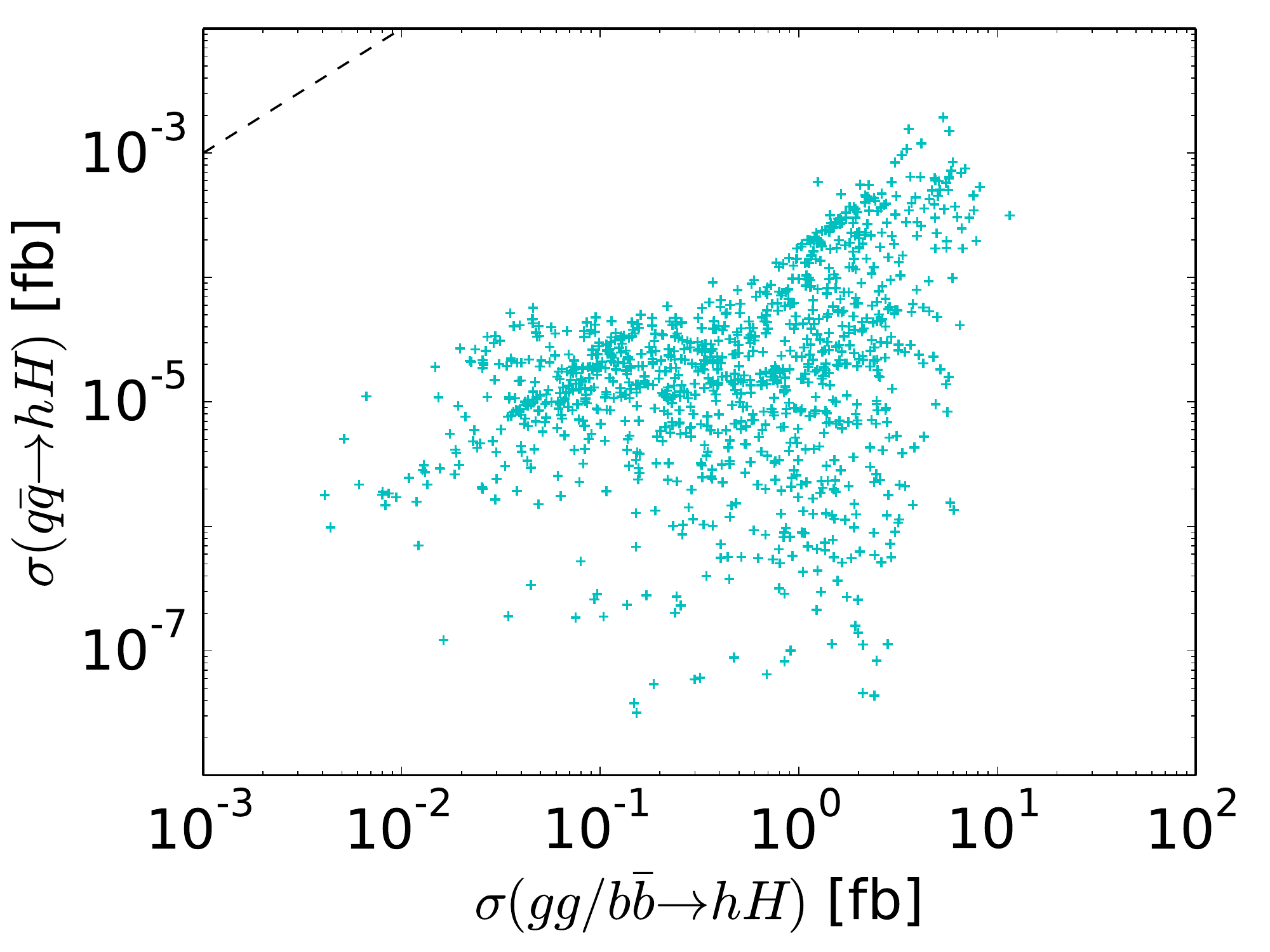}
\caption{Neutral 2BFSs for which $gg/b\bar{b}$ production by far dominates over $qq'$-initiated production. The dashed line indicates where the cross sections are of equal magnitude.}
\label{fig:neutral2bfsgg}
\end{center}
\end{figure}

\subsection{Neutral final states}

The neutral final states may be produced by $q\bar q$-induced processes as well as via loop-induced processes initiated by a pair of gluons. The cross sections for the $2\to 2$ neutral processes are shown in Figs.~\ref{fig:neutral2bfs} and \ref{fig:neutral2bfsgg} as a comparison between $q\bar{q}$ and $gg/b\bar{b}$ production. We find that, for $H^+H^-$, $h A$ and $H A$ final states, the $q\bar{q}$ cross sections can all exceed 10\,fb and, for some regions of the parameter space, dominate the combined $gg+b\bar{b}$ production, as shown in Fig.~\ref{fig:neutral2bfs}. While the remaining neutral 2BFSs, namely $hh$, $HH$, $AA$ and $hH$, have EW cross sections unlikely to be relevant at the LHC, their $gg/b\bar{b}$ production can be significant, as seen in Fig.~\ref{fig:neutral2bfsgg}, so these are still the more useful modes.

\begin{figure}[t!]
\includegraphics[width=0.33\textwidth]{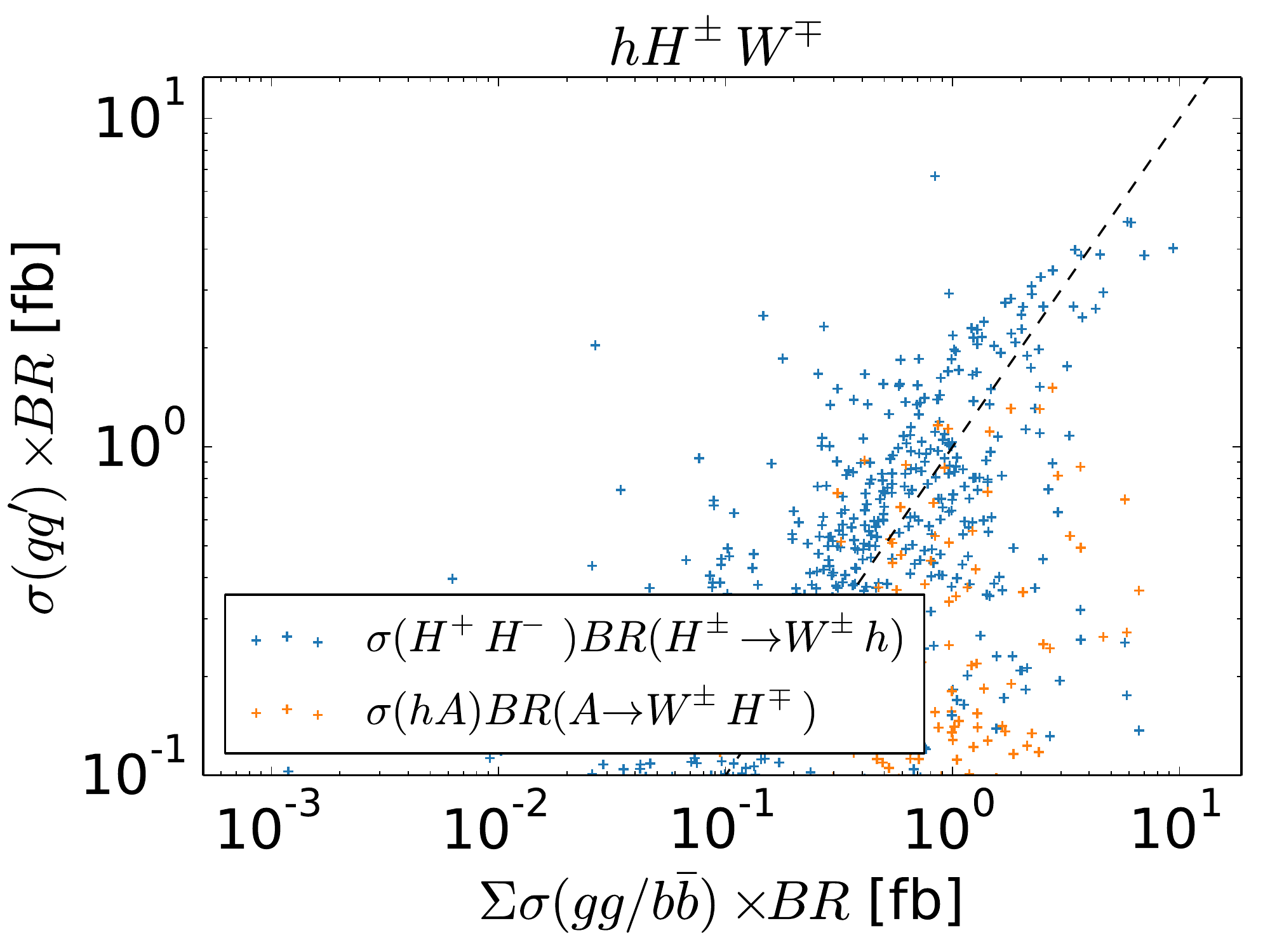}
\includegraphics[width=0.33\textwidth]{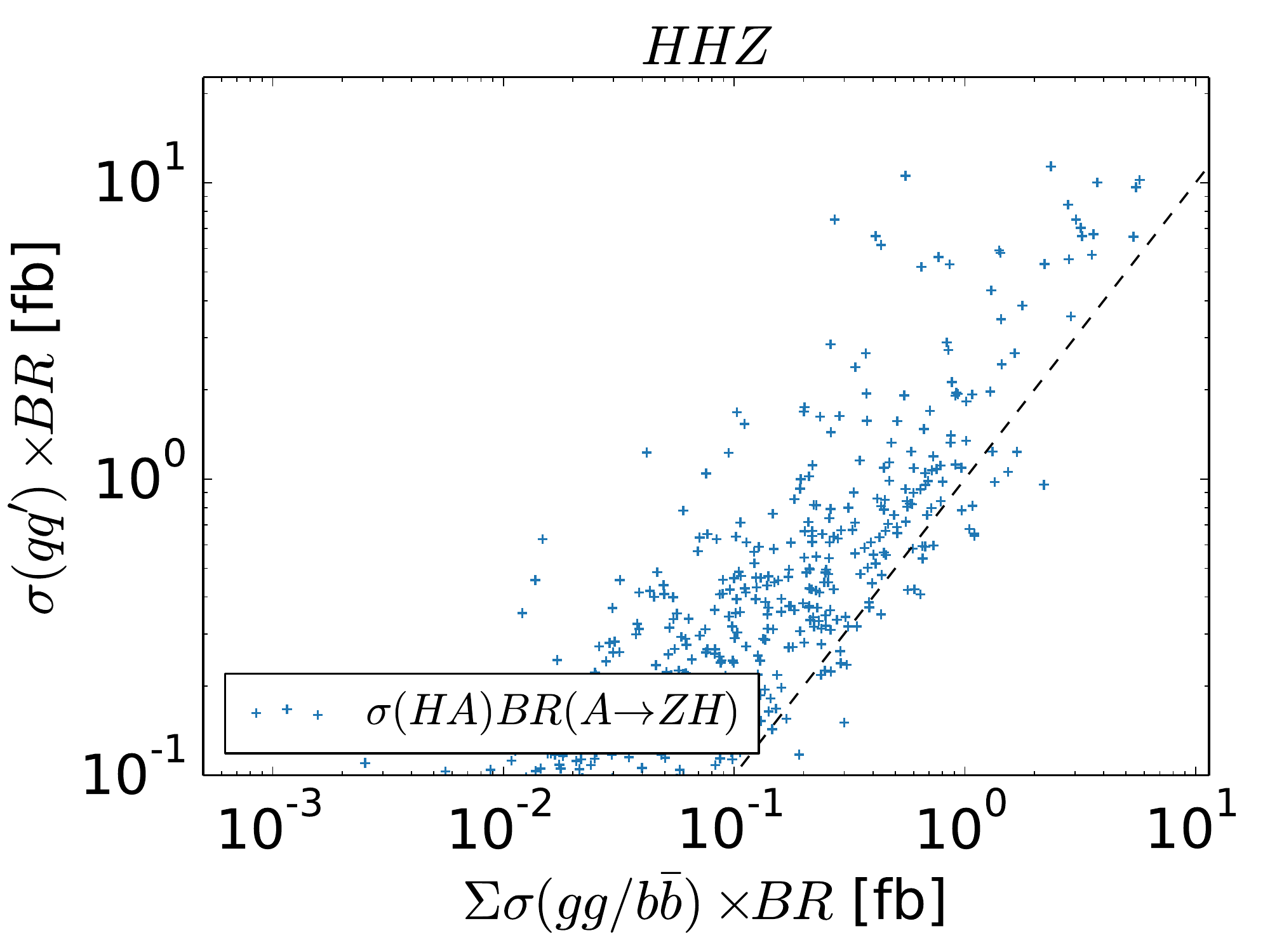}
\includegraphics[width=0.33\textwidth]{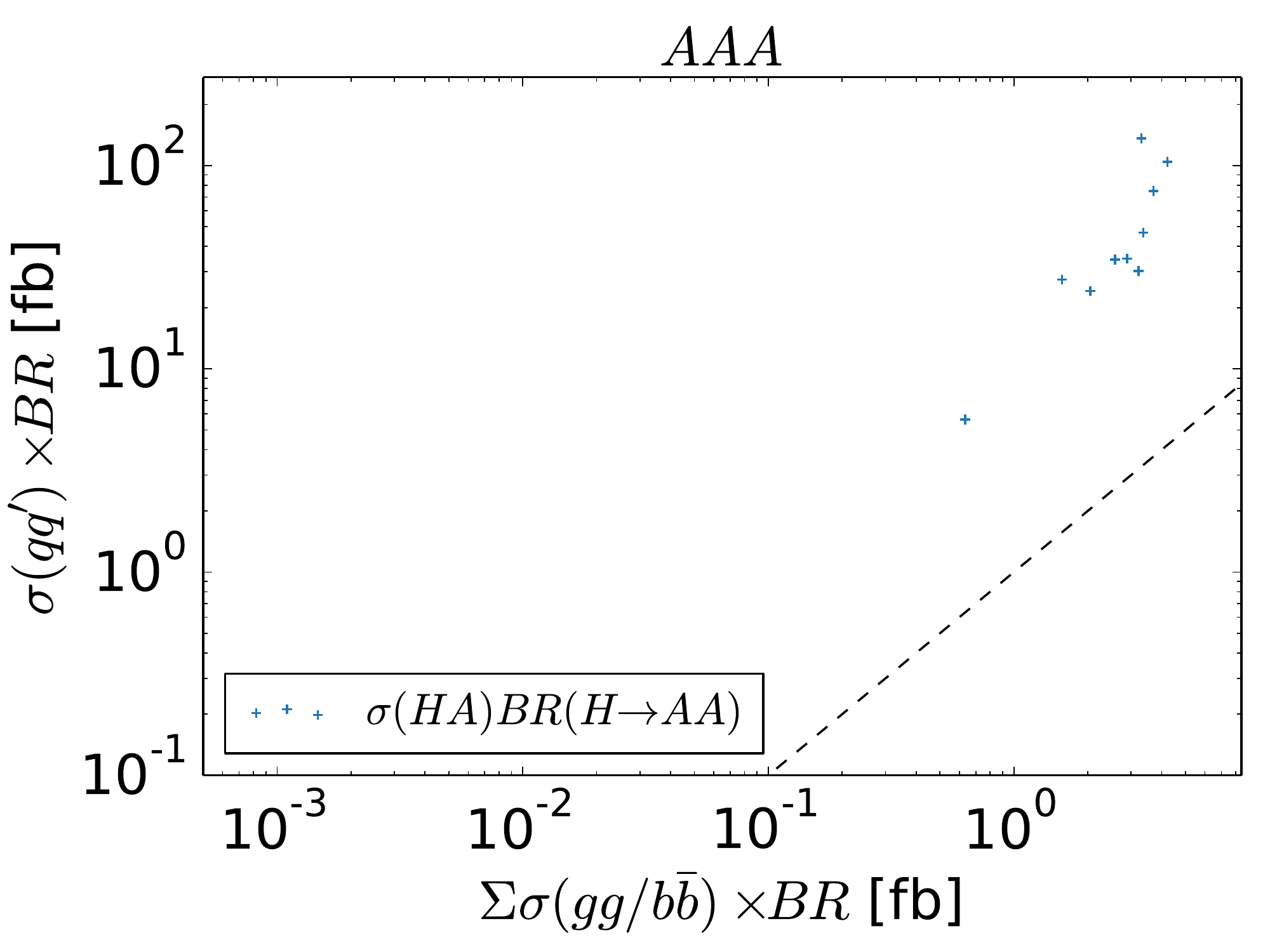}
\includegraphics[width=0.33\textwidth]{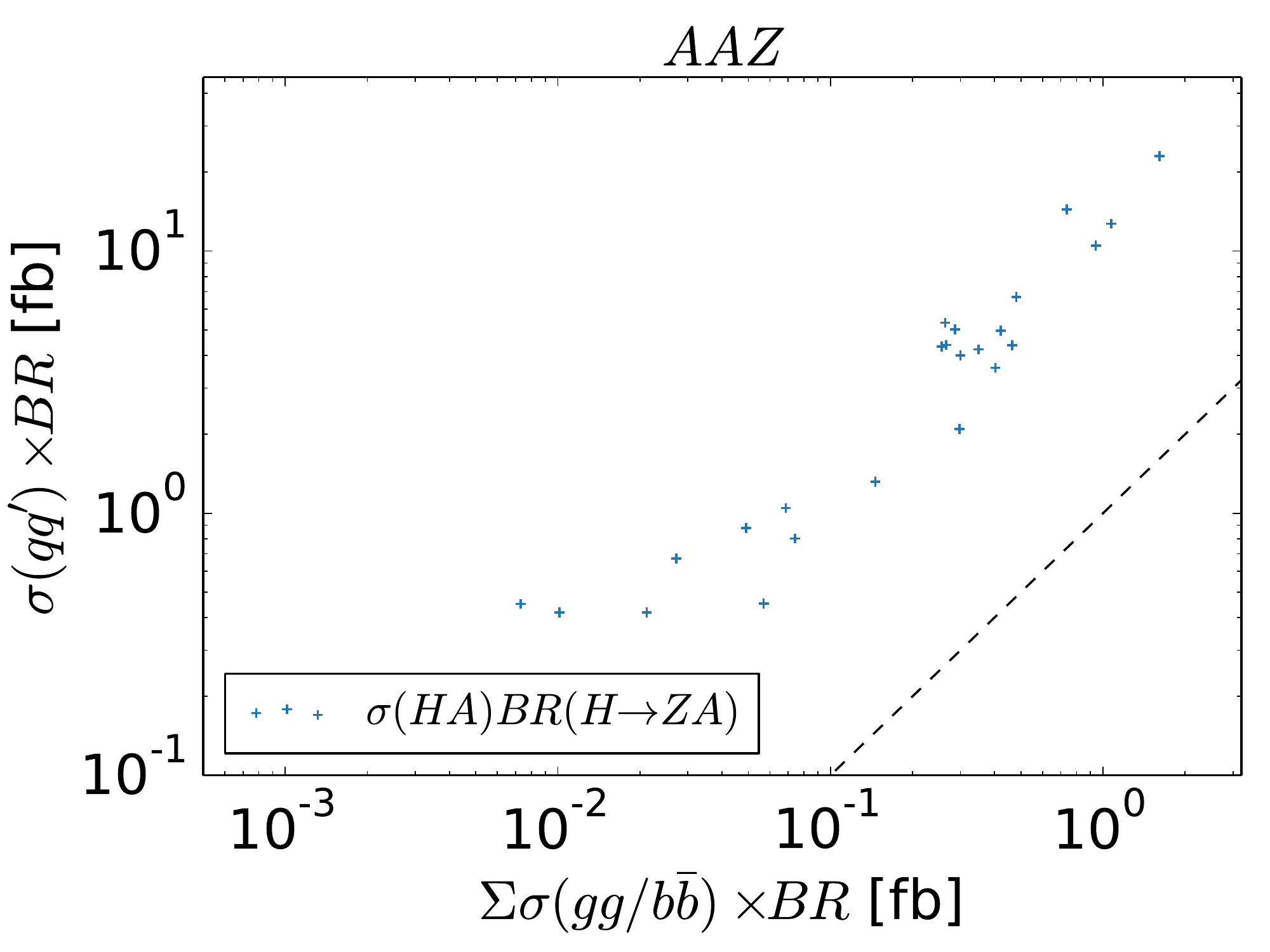}
\includegraphics[width=0.33\textwidth]{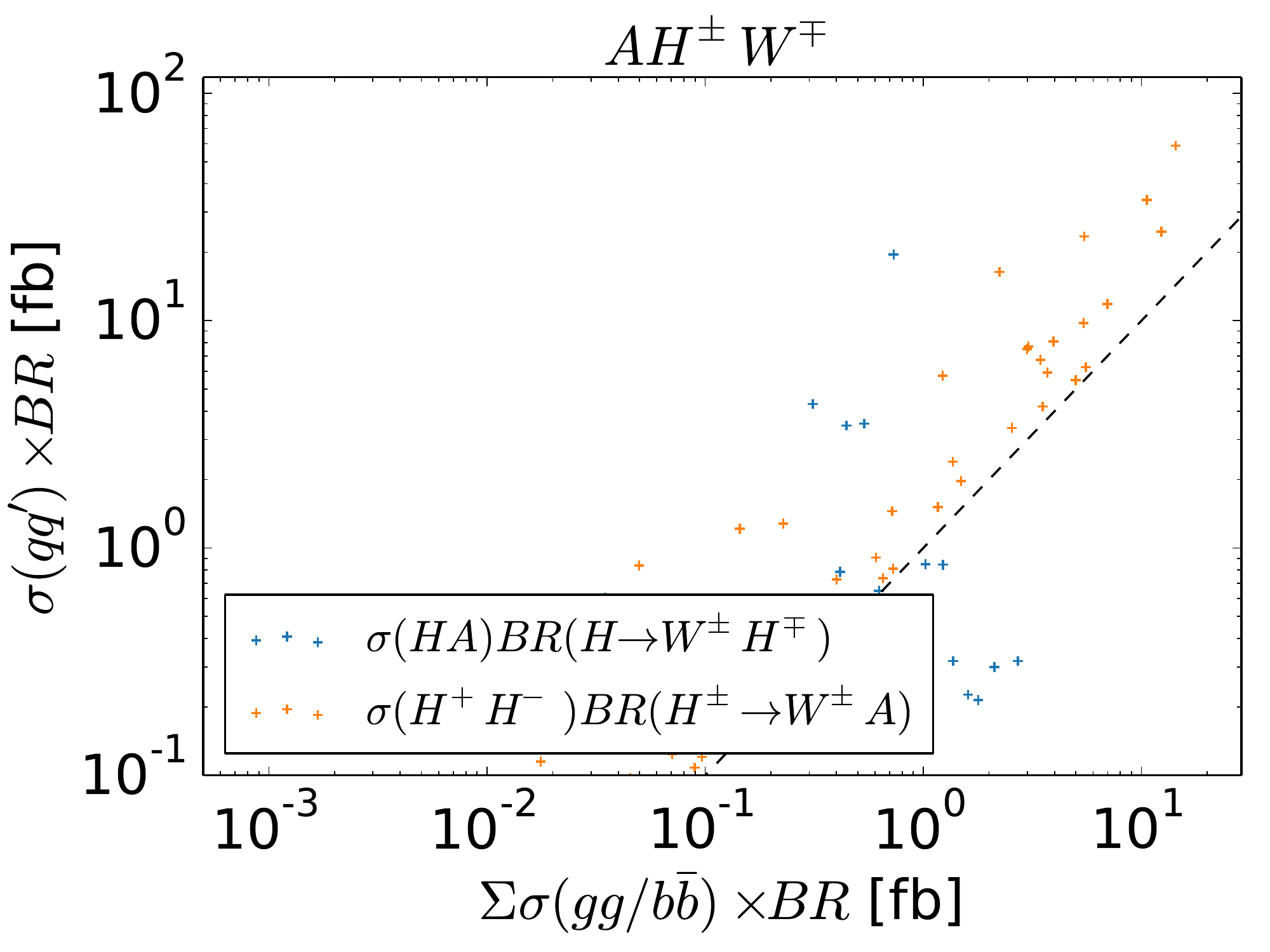}
\includegraphics[width=0.33\textwidth]{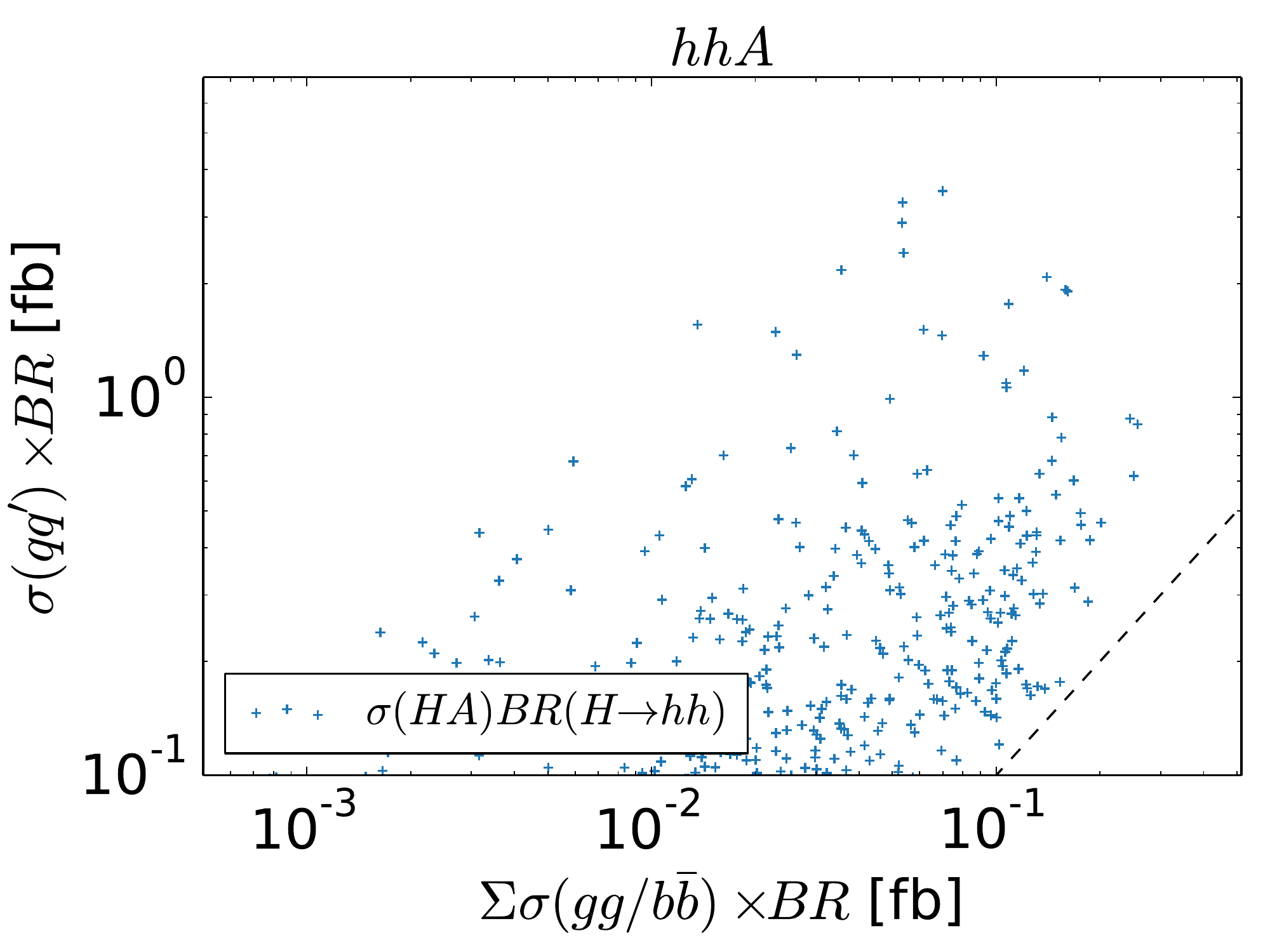}
\includegraphics[width=0.33\textwidth]{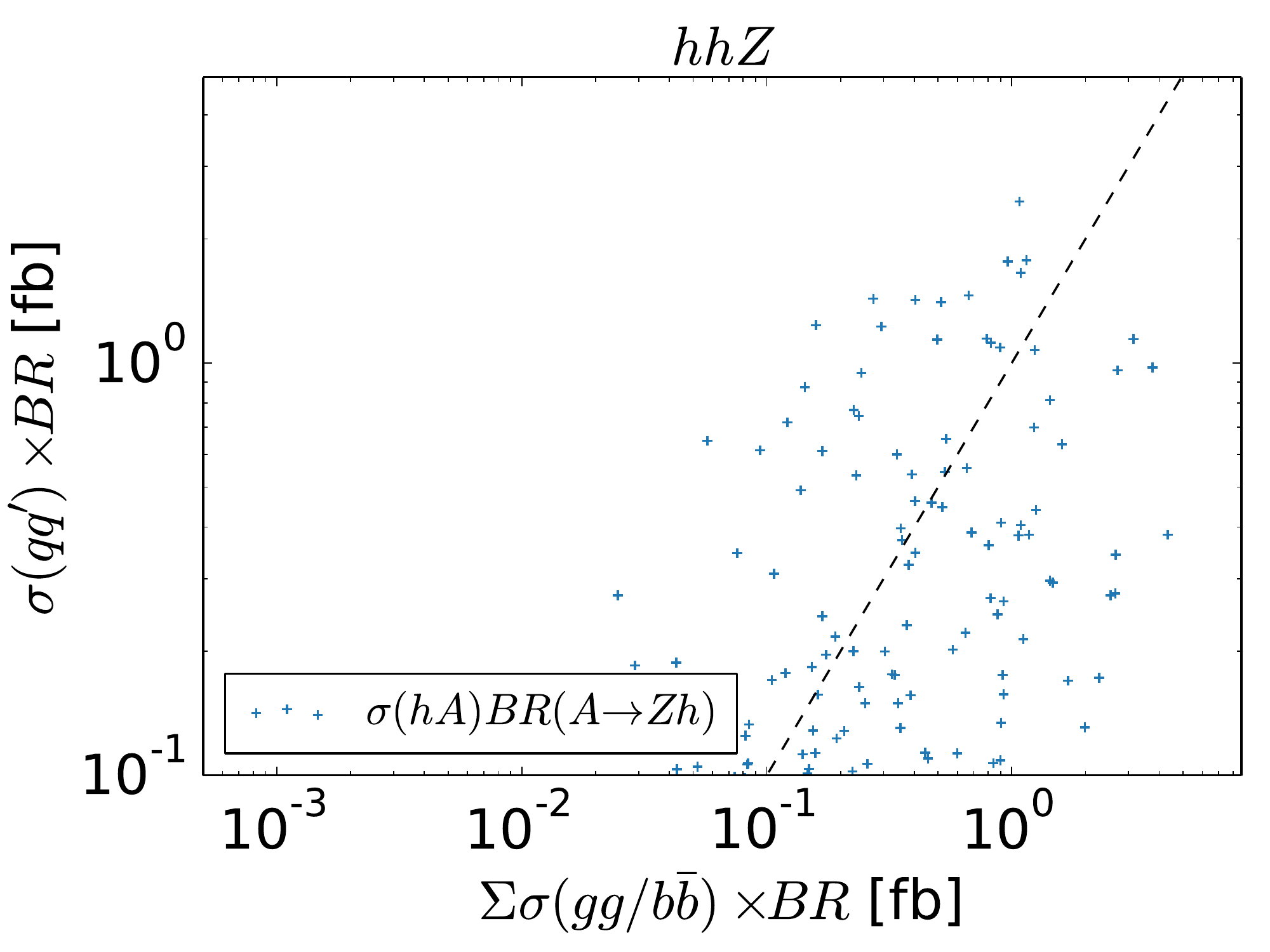}
\includegraphics[width=0.33\textwidth]{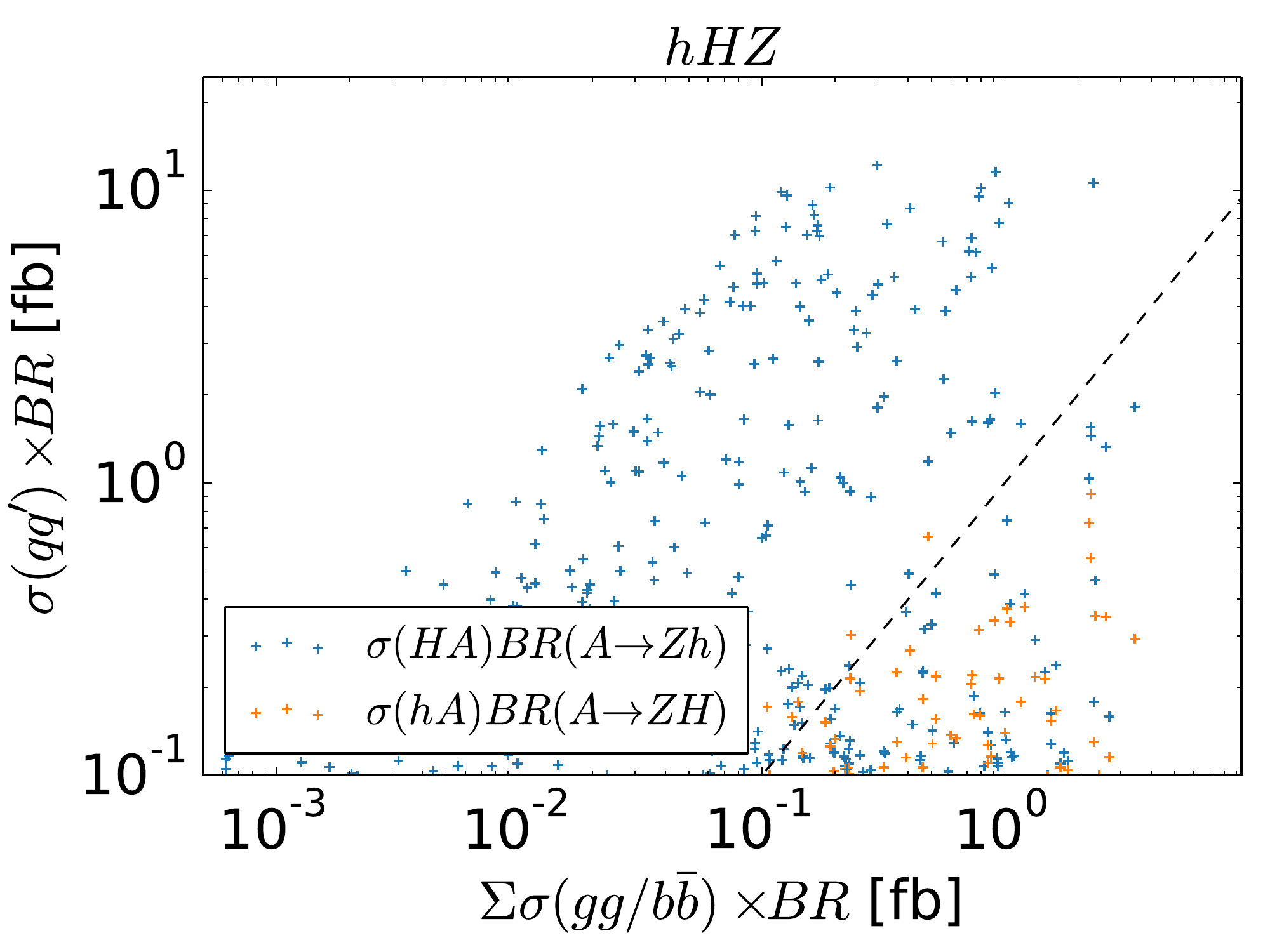}
\includegraphics[width=0.33\textwidth]{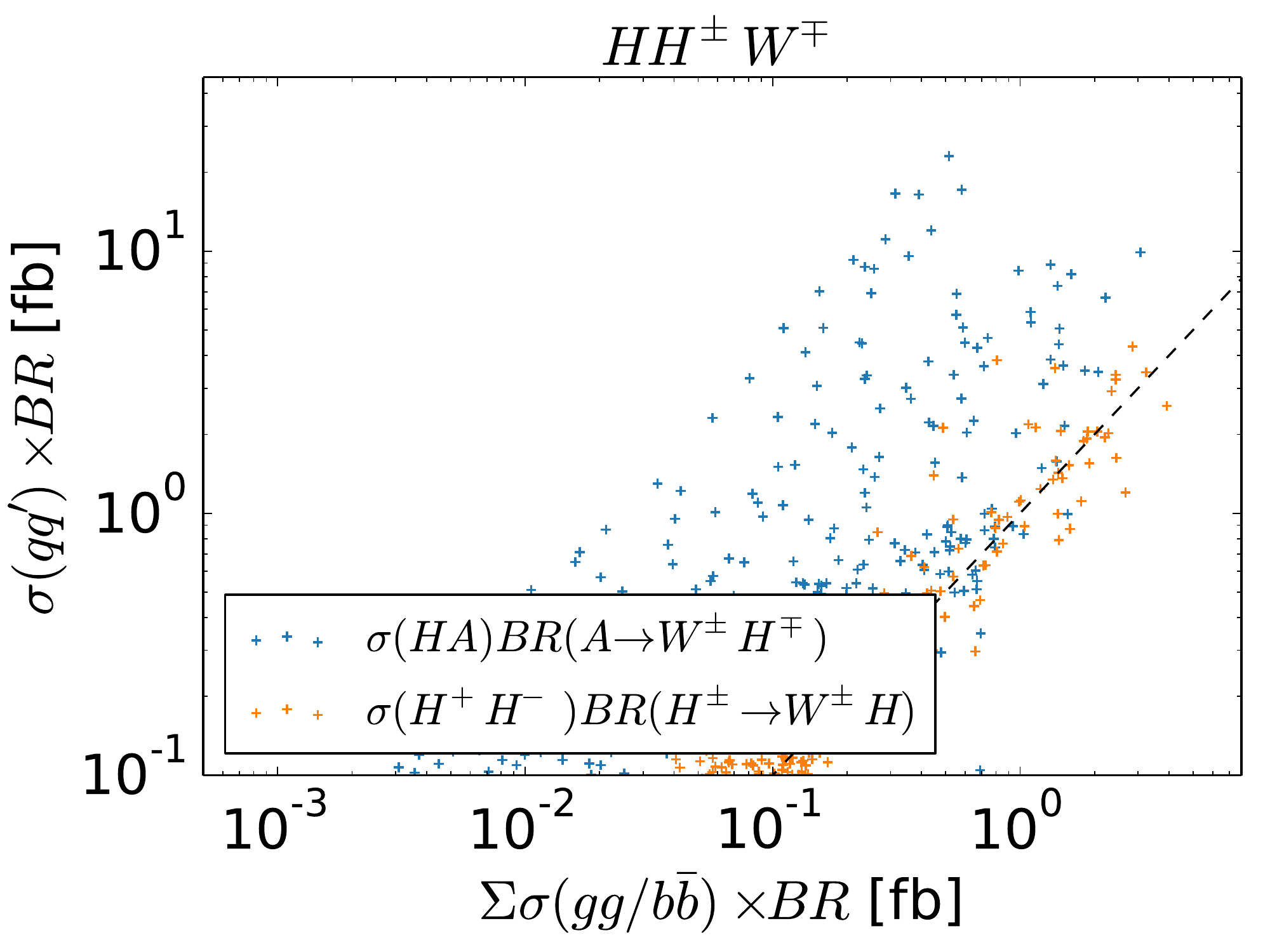}
\caption{Comparison of the cross sections for the $qq'$-initiated subprocesses and their $gg/bb$-initiated counterparts, for selected neutral 3BFSs. The dashed line indicates where the cross sections are of equal magnitude.}
\label{fig:neutral3bfs}
\end{figure}

For the neutral 2BFSs too we consider the possibility of one of the Higgs bosons decaying, and the resulting 3BFSs for which $q\bar{q}$ cross sections exceed 1\,fb are shown in Fig.~\ref{fig:neutral3bfs} and Tab.~\ref{tab:maxcx}. Again, we see some cross sections dominated by $q\bar{q}$ production. Here too all of the possible Higgs-to-Higgs decays are included, apart from $H\to H^+ H^-$, which is not kinematically available to any of our points. As with the charged 3BFSs, plots involving $H^\pm\to W^\pm A$, $H\to A A$, or $H\to A Z$ are sparsely populated, since these decays are only allowed for a small number of scanned points. 

\begin{table}[t!]
\centering\begin{tabular}{l|llc|llc|cc}
  & \multicolumn{2}{c}{Process 1} &  & \multicolumn{2}{c}{Process 2} &  & \\
3BFS  & 2BFS & BR & $\sigma_{qq'}^\textrm{max}$ & 2BFS & BR & $\sigma_{qq'}^\textrm{max}$ & $\sigma_{gg/bb}^\textrm{max}$  \\\hline

$AA\wpm$ & $$$A H^\pm$& $(H^\pm  \to  W^\pm  A)$ & 322 & & &  &  $-$   \\
$H^\pm H^\pm \wpm$ & $$$H H^\pm$& $(H \to  W^\pm  H^\mp )$ & 103 & $$$A H^\pm$ & $(A \to  W^\pm  H^\mp )$ & 94 &  $-$ \\
$AAH^\pm $ & $$$H H^\pm$& $(H \to  A A)$ & 95 & & &  &  $-$   \\
$HA\wpm$ & $$$H H^\pm$& $(H^\pm  \to  W^\pm  A)$ & 91 & $$$A H^\pm$& $(H^\pm  \to  W^\pm  H)$ & 12 &  $-$   \\
$hH^\pm Z$ & $$$A H^\pm$& $(A \to  Z h)$ & 22 & & &  &  $-$   \\
$HH\wpm$ & $$$H H^\pm$& $(H^\pm  \to  W^\pm  H)$ & 18 & & &  &  $-$   \\
$hH\wpm$ & $$$H H^\pm$& $(H^\pm  \to  W^\pm  h)$ & 16 & $$$h H^\pm$& $(H^\pm  \to  W^\pm  H)$ & 1 &  $-$   \\
$hA\wpm$ & $$$A H^\pm$& $(H^\pm  \to  W^\pm  h)$ & 15 & $$$h H^\pm$& $(H^\pm  \to  W^\pm  A)$ & 6 &  $-$  \\
$AH^\pm Z$ & $$$H H^\pm$& $(H \to  Z A)$ & 13 & & &  &  $-$  \\
$HH^\pm Z$ & $$$A H^\pm$& $(A \to  Z H)$ & 8 & & &  &  $-$   \\
$hhH^\pm $ & $$$H H^\pm$& $(H \to  h h)$ & 7 & & &  &  $-$  \\
$hh\wpm$ & $$$h H^\pm$& $(H^\pm  \to  W^\pm  h)$ & 2 & & &  &  $-$ \\ \hline
$AAA$ & $$$H A$& $(H \to  A A)$ & 135 & & &  &  4  \\
$AH^\pm \wpm$ & $$$H^+ H^-$& $(H^\pm  \to  W^\pm  A)$ & 58 & $$$H A$& $(H \to  W^\pm  H^\mp )$ & 19 &  14   \\
$HH^\pm \wpm$ & $$$H A$& $(A \to  W^\pm  H^\mp )$ & 23 & $$$H^+ H^-$& $(H^\pm  \to  W^\pm  H)$ & 4 &  3  \\
$AAZ$ & $$$H A$& $(H \to  Z A)$ & 23 & & &  &  1  \\
$hHZ$ & $$$H A$& $(A \to  Z h)$ & 12 & & &  &  4   \\
$HHZ$ & $$$H A$& $(A \to  Z H)$ & 11 & & &  &  5  \\
$hH^\pm \wpm$ & $$$H^+ H^-$& $(H^\pm  \to  W^\pm  h)$ & 6 & $$$h A$& $(A \to  W^\pm  H^\mp )$ & 1 &  9 \\
$hhA$ & $$$H A$& $(H \to  h h)$ & 3 & & &  & 0.3 \\
$hhZ$ & $$$h A$& $(A \to  Z h)$ & 2 & & &  &  4  \\
\end{tabular}
\caption{Maximum cross sections for each process, in fb.  Only cross sections above 1 fb are included.}
\label{tab:maxcx}
\end{table}

\subsection{Higgs boson couplings from multi-Higgs states at the LHC}
Evidently, based on our results so far, several different processes and final states could potentially be observed at the LHC, thus serving as probes of the various couplings appearing in the 2HDM Lagrangian. In Tab.~\ref{tab:di-higgs} we have listed the ten triple-Higgs couplings (a -- h) and the six (pseudo)scalar-gauge couplings (i -- n) that appear in the 2HDM Lagrangian (assuming minimal flavour violation) row-wise and all the possible di-Higgs 2BFS combinations column-wise. If a coupling can potentially enter the secondary vertex of both $gg/b\bar{b}$- and $q\bar{q}$-initiated $s$-channel production processes of a given 2BFS at the LHC, the corresponding cell is checked. 

\begin{figure}[t!]
 \centering\includegraphics[scale=0.9]{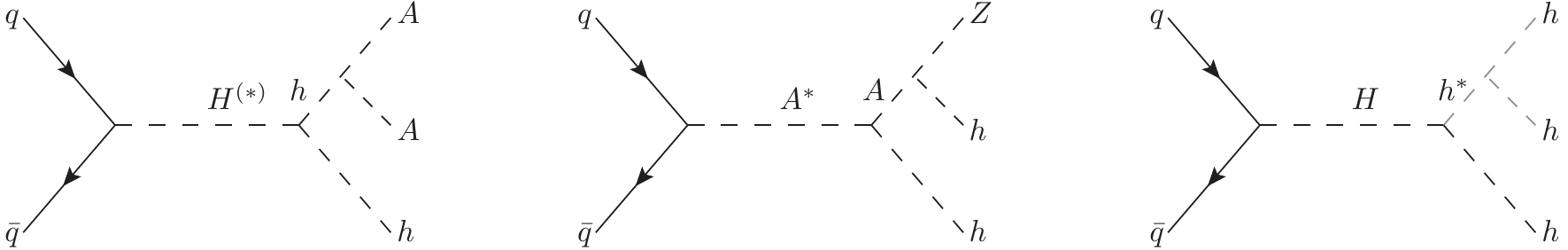}
 \caption{\label{fig:graphs} Examples of $s$-channel diagrams considered for the production of three-body final states. Processes like the one on the right are not taken into account in the scatter plots shown above, as two of the three final state particles result from an incoming Higgs state that is necessarily off-shell. Thus the corresponding cross sections cannot be calculated as $\sigma(gg/b\bar{b}/q\bar{q}^{(\prime)} \to h_ih_x)$*BR($h_x\to h_j+h_k/V_k$). Such 3BFSs have therefore been typeset in grey colour in Tab.~\ref{tab:hig-cplg}.}
\end{figure}

In Tab.~\ref{tab:hig-cplg} we similarly show possible 3BFSs, comprising of at least two Higgs bosons and at most one gauge boson, that can originate from the 2BFS at the top of a column. For a given 3BFS, the coupling at the start of the corresponding row appears in, instead of the secondary vertex in the production process of its parent 2BFS, the tertiary vertex between one of the two incoming Higgs bosons and an outgoing Higgs+Higgs/gauge state. A 3BFS has a `$^*$' next to it if the incoming Higgs state is necessarily off-shell, i.e., if its mass, $m_x$, is smaller than the sum of the masses, $m_j + m_k$, of the two outgoing particles. In such a case, the cross section for the corresponding process cannot be evaluated in the $\sigma(gg/b\bar{b}/q\bar{q}^{(\prime)} \to h_ih_x)\times$ BR($h_x\to h_j+h_k/V_k$) approach adopted here, and it therefore does not contribute to the cumulative cross section shown for a given 3BFS in the scatter plots in the previous sections.\footnote{We note that virtual exchanges could also be potentially relevant for the cases where $m_x > m_j + m_k$, especially just above threshold. However, we again expect that the cross sections highlighted here will receive relatively small corrections from such contributions owing to the narrow widths of the intermediate states.} The rightmost graph in Fig.~\ref{fig:graphs} illustrates this scenario. In both the tables, charged final states are typeset in bold and a box appears around those for which the total ($q\bar{q}$) production cross section can be larger than 1\,fb, while a box around a neutral final state indicates that the cross section for $q\bar{q}$ production can exceed that for $gg/b\bar{b}$ production (for certain parameter space configurations).  

\begin{table}[t!]
\resizebox{6.5in}{!}{%
\centering\begin{tabular}{|c|c|c|c|c|c|c|c|c|c|c|}
  \hline
 Coupling & 1.\,$hh$ & 2.\,$HH$ & 3.\,$AA$ & 4.\,\bbtcr{$\hp\hm$} & 5.\,$hH$ & 6.\,\bbtcr{$hA$} & 7.\,\boldit{h}\hpmb & 8.\,\bbtcr{$HA$} & 9.\,\boldit{H}\hpmb & 10.\,\boldit{A}\hpmb \\ \hline
a.\,$\lambda_{hhh}$ & \cm & & & & & & & & & \\ \hline

b.\,$\lambda_{hhH}$ & \cm & & & & \cm & & & & & \\ \hline

c.\,$\lambda_{hHH}$ & & \cm & & & \cm & & & & & \\ \hline

d.\,$\lambda_{hAA}$ & & & \cm & & & \cm & & & & \\ \hline

e.\,$\lambda_{h\hp\hm}$ & & & & \cm & & & \cm & & & \\ \hline

f.\,$\lambda_{HHH}$ & & \cm & & & & & & & & \\ \hline

g.\,$\lambda_{HAA}$ & & & \cm & & & & & \cm & & \\ \hline
                              
h.\,$\lambda_{H\hp\hm}$ & & & & \cm & & & & & \cm & \\ \hline

i.\,$\lambda_{hAZ}$ & & & & & & \cm & & & & \\ \hline
               
j.\,$\lambda_{HAZ}$ & & & & & & & & \cm & & \\ \hline

k.\,$\lambda_{\hp\hm Z}$ & & & & \cm & & & & & & \\ \hline

l.\,$\lambda_{h\hp\wmi}$ & & & & & & & \cm & & & \\ \hline

m.\,$\lambda_{H\hp\wmi}$ & & & & & & & & & \cm & \\ \hline       
                                                                       
n.\,$\lambda_{A\hp\wmi}$ & & & & & & & & & & \cm \\ \hline
\end{tabular}
    }
\caption{\label{tab:di-higgs} The ten 2BFS combinations available in the 2HDM. Charged 2BFSs, which can only be $q\bar{q}^{(\prime)}$-produced, are typeset in bold in the top row, while a box around a neutral 2BFS implies that the cross section for its production from $q\bar{q}^{(\prime)}$-initiated processes can exceed that from $gg/bb$-initiated processes. A \cm\ appears in a cell if the coupling at the start of the corresponding row may enter the $s$-channel production of the given 2BFS.}
\end{table}

\begin{table}[t!]
\resizebox{6.5in}{!}{%
\centering\begin{tabular}{|c|c|c|c|c|c|c|c|c|c|c|}
  \hline
 Coupling & 1.\,$hh$ & 2.\,$HH$ & 3.\,$AA$ & 4.\,\bbtcr{$\hp\hm$} & 5.\,$hH$ & 6.\,\bbtcr{$hA$} & 7.\,\boldit{h}\hpmb & 8.\,\bbtcr{$HA$} & 9.\,\boldit{H}\hpmb & 10.\,\boldit{A}\hpmb \\ \hline
 
a.\,$\lambda_{hhh}$ & $(hhh)^*$ & & & & $(hhH)^*$ & $(hhA)^*$ & $(hh\hpm)^*$ & & & \\ \hline

b.\,$\lambda_{hhH}$ & & $hhH$ & & & $hhh$ & & & \rbtcb{$hhA$} & \bbtcr{\boldit{hh}\hpmb} & \\ \hline

\mr{2}{c.\,$\lambda_{hHH}$} & & \mr{2}{$(hHH)^*$} & & & $(hhH)^*$ & & & \mr{2}{$(hHA)^*$} & \mr{2}{$(hH\hpm)^*$} & \\
                            & &                   & & & $h\hp\hm$ & & &                   &                      & \\ \hline

\mr{2}{d.\,$\lambda_{hAA}$} & \mr{2}{$(hAA)$} & & \mr{2}{$(hAA)^*$} & \mr{2}{$(h\hp\hm)^*$} & \mr{2}{$HAA$} & $(hhA)^*$ & \mr{2}{$(AA\hpm)^*$} & \mr{2}{$(hHA)^*$} & & \\ 
                            &                 & &                   &                       &               & $AAA$                                                                                                                                                                                                 &                               &                   & &  \\ \hline

\mr{2}{e.\,$\lambda_{h\hp\hm}$} & \mr{2}{$h\hp\hm$} & & & \mr{2}{$(h\hp\hm)^*$} & \mr{2}{$H\hp\hm$} & \mr{2}{$A\hp\hm$} & $(hh\hpm)^*$ & & \mr{2}{$(hH\hpm)^*$} & \mr{2}{$(hA\hpm)^*$} \\ 
                                &                   & & &                       &                   &                   & \hpb\hmb\hpmb   & &                     &                      \\ \hline

f.\,$\lambda_{HHH}$ & & $(HHH)^*$ & & & $(hHH)^*$ & & & $(HHA)^*$ & $(HH\hpm)^*$ &  \\ \hline

\mr{2}{g.\,$\lambda_{HAA}$} & & \mr{2}{$HAA$} & \mr{2}{$(HAA)^*$} & & \mr{2}{$hAA$} & \mr{2}{$(hHA)^*$} & & $(HHA)^*$ & \mr{2}{\bbtcr{\boldit{AA}\hpmb}} & \mr{2}{\boldit{HA}\hpmb} \\ 
                            & &               &                   & &               &                   & & \rbtcb{$AAA$}  &                   &                          \\ \hline
                              
\mr{2}{h.\,$\lambda_{H\hp\hm}$} & & \mr{2}{$H\hp\hm$} & & \mr{2}{$(H\hp\hm)^*$} & & & \mr{2}{$(hH\hpm)^*$} & \mr{2}{$A\hp\hm$} & $(HH\hpm)^*$ & \mr{2}{$(HA\hpm)^*$}  \\ 
                                & &                   & &                       & & &                      &                                            &          \hpb\hmb\hpmb   &                       \\ \hline

\mr{2}{i.\,$\lambda_{hAZ}$} & \mr{2}{$hAZ$} & & \mr{2}{$hAZ$} & & \mr{2}{$HAZ$} & \rbtcb{$hhZ$} & \mr{2}{\boldit{A}\hpmb\boldit{Z}} & \mr{2}{\rbtcb{$hHZ$}} & & \mr{2}{\bbtcr{\boldit{h}\hpmb\boldit{Z}}} \\
                            &               & &               & &               &    $AAZ$      &                     &                                        & &                                           \\ \hline
               
\mr{2}{j.\,$\lambda_{HAZ}$} & & \mr{2}{$HAZ$} & \mr{2}{$HAZ$} & & \mr{2}{$hAZ$} & \mr{2}{\rbtcb{$hHZ$}} & & \rbtcb{$HHZ$} & \mr{2}{\bbtcr{\boldit{A}\hpmb\boldit{Z}}} & \mr{2}{\bbtcr{\boldit{H}\hpmb\boldit{Z}}} \\ 
                            & &               &               & &               &                       & &                                                 \rbtcb{$AAZ$} &                                           &                                           \\ \hline

k.\,$\lambda_{\hp\hm Z}$ & & & & $\hp\hm Z$ & & & & & & \\ \hline

\mr{2}{l.\,$\lambda_{h\hp\wmi}$} & \mr{2}{$h\hp\wmi$} & & & \mr{2}{\rbtcb{$h\hp\wmi$}} & \mr{2}{$H\hp\wmi$} & \rbtcb{$h\hp\wmi$} & \bbtcr{\boldit{hh}\wpmb} & & \mr{2}{\bbtcr{\boldit{hH}\wpmb}} & \mr{2}{\bbtcr{\boldit{hA}\wpmb}}  \\
                                 &                    & & &                            &                    &       $A\hp\wmi$         & \hpb\hmb\wpmb            & &                                  &                                                         \\ \hline

\mr{2}{m.\,$\lambda_{H\hp\wmi}$} & & \mr{2}{$H\hp\wmi$} & & \mr{2}{\rbtcb{$H\hp\wmi$}} & \mr{2}{$h\hp\wmi$} & & \mr{2}{\bbtcr{\boldit{hH}\wpmb}} & \rbtcb{$H\hp\wmi$} & \bbtcr{\boldit{HH}\wpmb} & \mr{2}{\bbtcr{\boldit{HA}\wpmb}}\\
                                 & &                    & &                            &                    & &       &                                 \rbtcb{$A\hp\wmi$}  & \bbtcr{\hpb\hmb\wpmb}    &                                 \\ \hline       
                                                                       
\mr{2}{n.\,$\lambda_{A\hp\wmi}$} & & & \mr{2}{$A\hp\wmi$} & \mr{2}{\rbtcb{$A\hp\wmi$}} & & & \mr{2}{\bbtcr{\boldit{hA}\wpmb}} & & \mr{2}{\bbtcr{\boldit{HA}\wpmb}} & \bbtcr{\boldit{AA}\wpmb} \\ 
                                 & & &                    &                            & & &                                &                            &                                  & \bbtcr{\hpb\hmb\wpmb} \\ \hline
\end{tabular}
    }
\caption{\label{tab:hig-cplg} 3BFSs that can result from the decay, via a vertex involving the coupling at the start of a given row, of one of the Higgs bosons in the 2BFS at the top of the column. Again, a charged 3BFS has a box around it if its total cross section can exceed 1\,fb, while a box around a neutral 3BFS indicates that its $q\bar{q}^{(\prime)}$ production can dominate over $gg/bb$ production. A `$^*$' next to a 3BFS implies that its cross section has not been calculated in this study. See text for more details.}
\end{table}

There are some important inferences that can be drawn from the table (note again that all the statements regarding the 3BFSs are valid only in the parameter space regions that satisfy $m_x > m_j + m_k$). One can notice many instances where a coupling appears in more relevant 3BFSs than 2BFSs. While a given 2BFS typically reflects contributions from several diagrams containing different couplings, the 3BFSs often arise from multiple initial 2BFSs, and the decays leading to 3BFSs reflect not only the relevant coupling, but also all other couplings and masses involved in determining the width of the decaying particle. A careful kinematical selection of events might help disentangle (some of) these couplings from each other, and complementary analyses of the two types of states can greatly enhance the potential of the LHC to probe such couplings.

While only $q\bar{q}^{(\prime)}$-production is available at leading order for charged 3BFSs, it is clearly the preferred mode also for several neutral 3BFSs, especially those involving the $\lambda_{hAZ}$, $\lambda_{HAZ}$ and $\lambda_{h\hp \wmi}$, $\lambda_{H\hp\wmi}$ couplings. Additionally, we see that all of the charged 3BFSs that include a $\wpm$ can have a cross section in excess of 1\,fb, which is a consequence of the cross section for the $H\hpm$ and $A\hpm$ 2BFSs reaching up to 100\,fb, as noted in Fig.\,\ref{fig:charged2bfs} earlier. As a result, $q\bar{q}^{(\prime)}$-production of the relevant 3BFSs, if observed, could prove crucial for pinning down the $\lambda_{h\hp \wmi}$, $\lambda_{H\hp\wmi}$ and $\lambda_{A\hp\wmi}$ couplings at the LHC. 

\subsection{The triple-Higgs couplings}

\begin{figure}[t!]
\includegraphics[angle=0,width=0.5\textwidth]{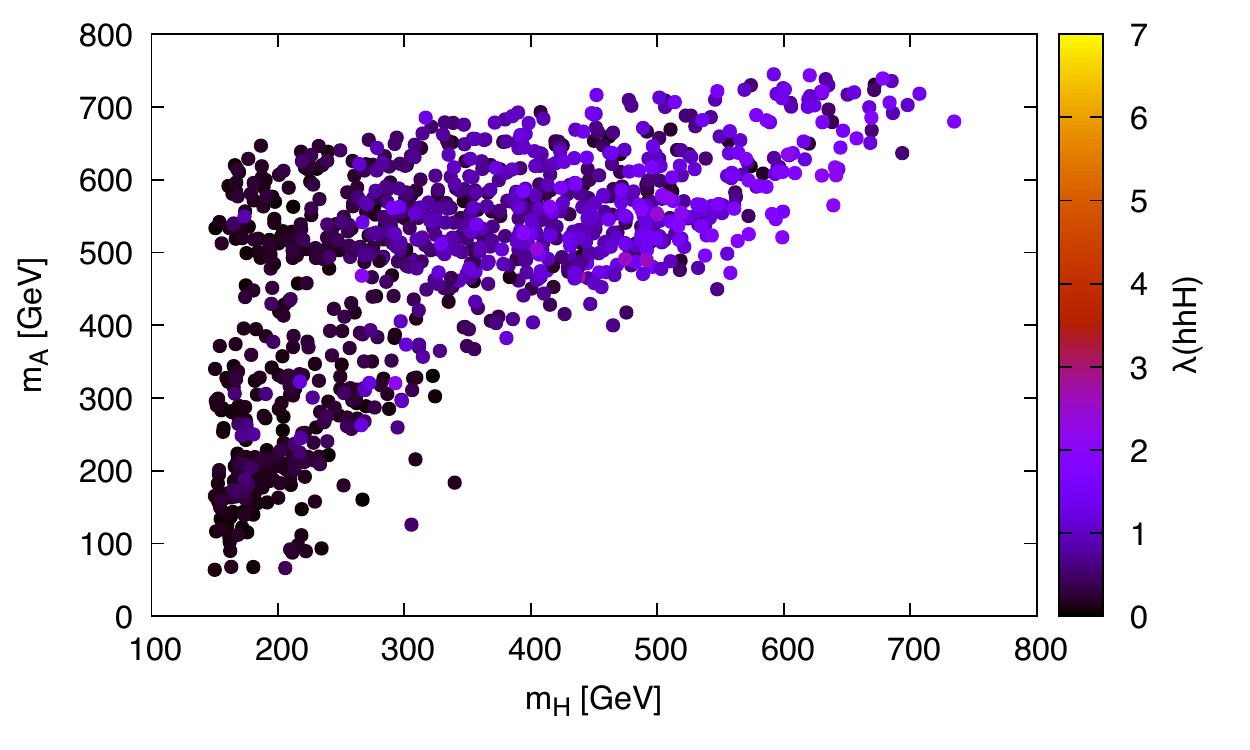}
\includegraphics[angle=0,width=0.5\textwidth]{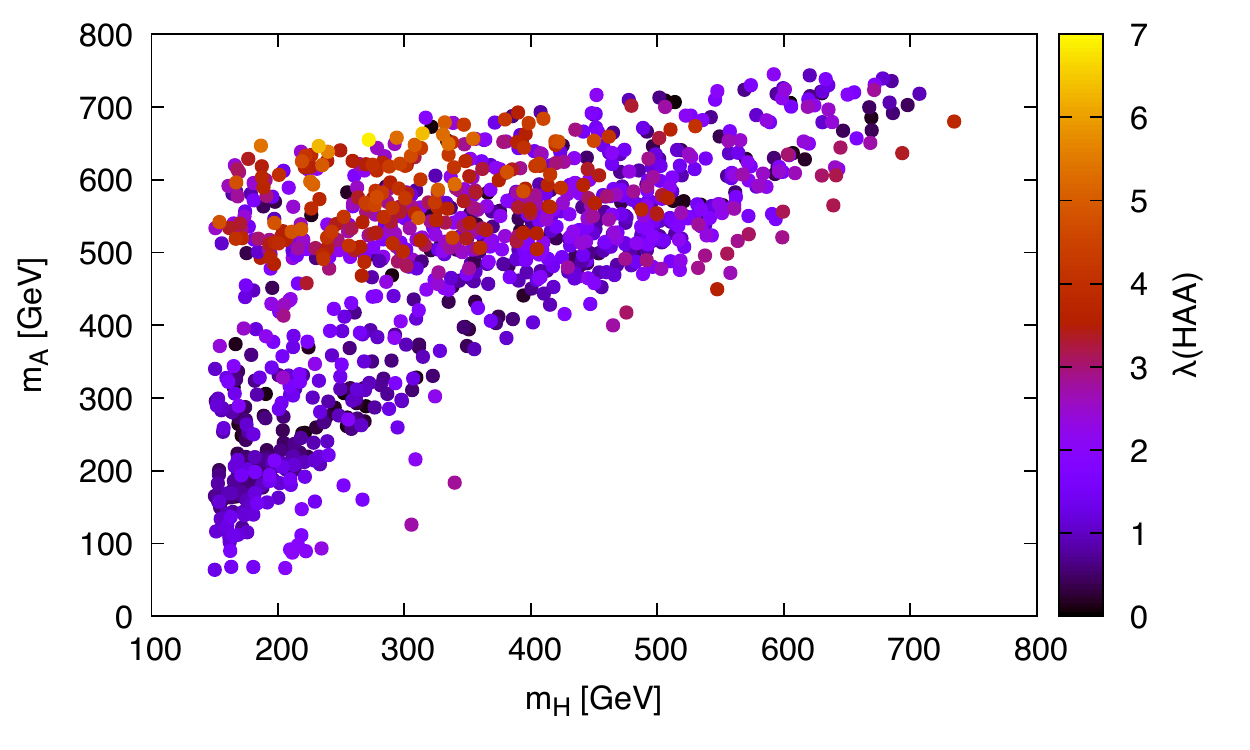}
\caption{The triple-Higgs couplings $\lambda_{hhH}$ and  $\lambda_{HAA}$, in units of the value of the Higgs triple self-coupling in the SM, shown by the color scale in the plane of the masses of the heavy CP-even and CP-odd
 neutral scalars.}
\label{fig:triple2}
\end{figure}

\begin{figure}[t!]
\includegraphics[angle=0,width=0.5\textwidth]{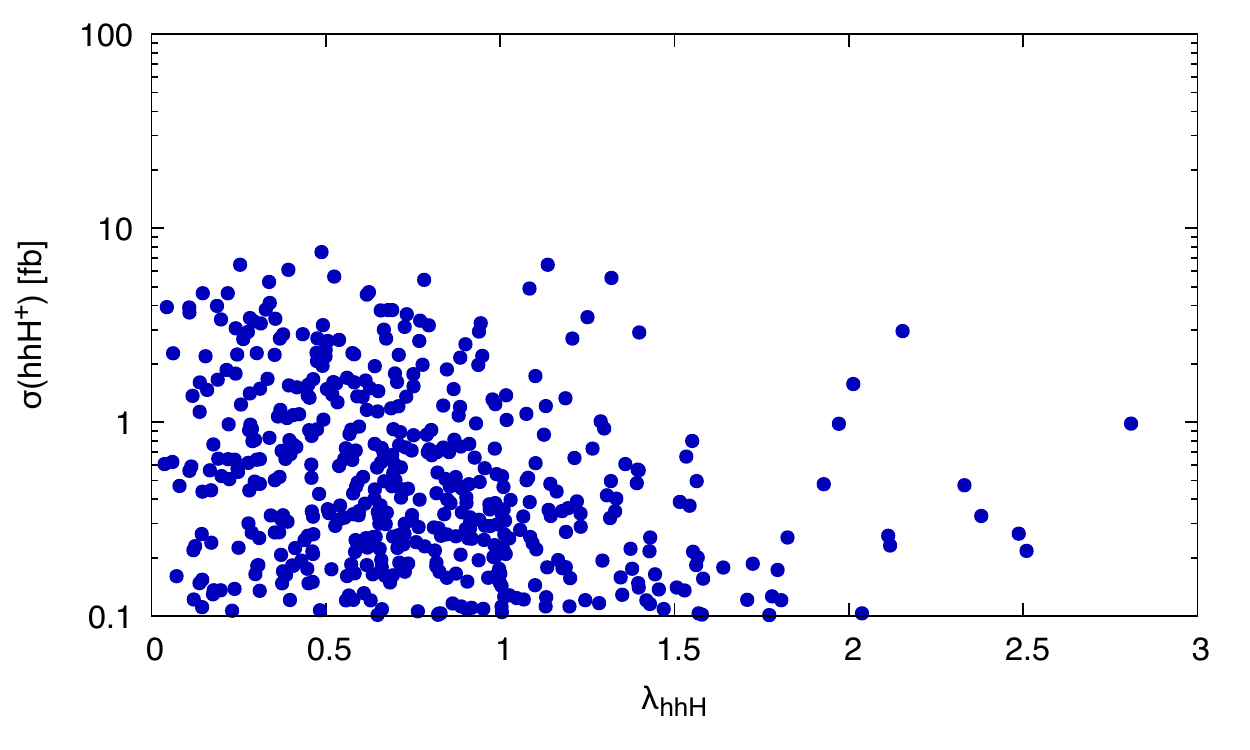}
\includegraphics[angle=0,width=0.5\textwidth]{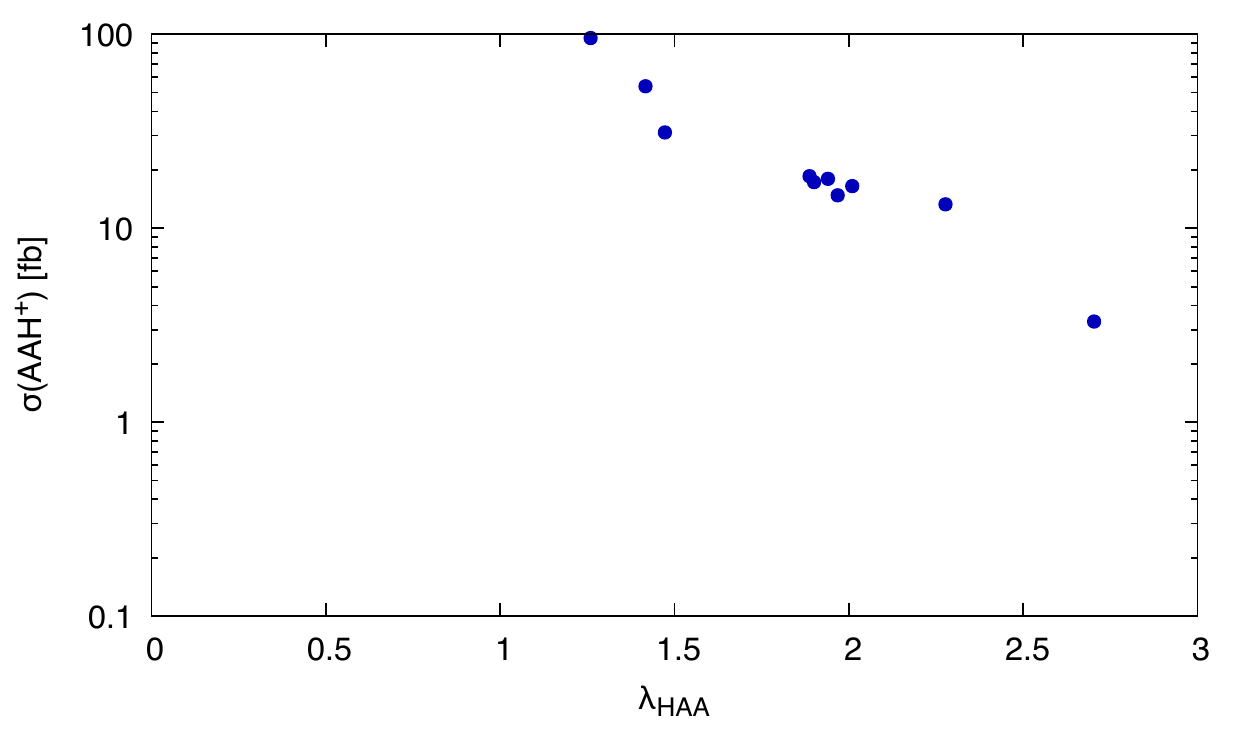} 
\caption{Cross-sections $\sigma(hhH^{\pm})$ and $\sigma(AAH^{\pm})$ plotted against the triple-Higgs couplings $\lambda_{hhH}$ and $\lambda_{HAA}$, respectively, with the couplings plotted in units of the value of the Higgs triple self-coupling in the SM, $3m_h^2/v$. The cross sections are the same as those plotted in Fig.~\protect\ref{fig:charged3bfs}.}
\label{fig:triple3}
\end{figure}

Of particular relevance for disentangling the underlying Higgs dynamics are the triple-Higgs couplings. In Tab.~\ref{tab:hig-cplg}, rows b and g, we see that the couplings $\lambda_{hhH}$ and $\lambda_{HAA}$ enter, respectively, in processes for which EW production dominates for neutral 3BFSs $hhA$ and $AAA$, and at the same time, also in EW processes giving substantial cross sections for charged 3BFSs $hhH^{\pm}$ and $AAH^{\pm}$. In order to give an impression of the possible sizes of the $\lambda_{hhH}$ and $\lambda_{HAA}$ couplings, the colour heat map in Fig.~\ref{fig:triple2} shows them in units of the SM-like Higgs self-coupling $\lambda_{hhh}$, as functions of the neutral scalar masses $m_H$ and $m_A$. We further show the cross sections for $hhH^{\pm}$ and $AAH^{\pm}$ production as functions of $\lambda_{hhH}$ and $\lambda_{HAA}$, respectively, in Fig.~\ref{fig:triple3}. 
 
The $\lambda_{hhH}$ and $\lambda_{HAA}$ couplings range from essentially zero up to several times larger than the Higgs self-coupling in the SM. The $\lambda_{HAA}$ coupling in particular can be sizeable, and may lead to a large $\sigma(AAH^{\pm})$, although a relatively small portion of the scanned parameter space lies above the threshold for this process, as was previously also noted in the central panel of the lowermost row in Fig.~\ref{fig:charged3bfs}. On the other hand, the production of $hhH^{\pm}$, which is sensitive to the $\lambda_{hhH}$ coupling, is kinematically allowed over a much larger portion of our scanned parameter space. While the cross section for this 3BFS is generally smaller than $\sigma(AAH^{\pm})$, it can still reach upto 10\,fb. 

\section{Conclusions}\label{sec:summa}
In order to fully establish the EWSB mechanism, whether in the SM theory or beyond it, a full reconstruction of the Higgs potential is required. This entails measuring experimentally the triple-Higgs couplings, which can only be achieved if scattering processes yielding two or more Higgs bosons can be isolated in the detector. Historically, most studies of these couplings have exploited production modes that are enhanced in the hadronic environment of the LHC, primarily gluon-gluon fusion. Such studies have covered both the SM as well as extended Higgs sectors, chiefly 2HDMs, with and without Supersymmetry. In such beyond-the-SM scenarios, couplings of the Higgs bosons to $b$-(anti)quarks can be enlarged with respect to the SM case, so that $b\bar b$-induced production can be relevant in onsetting final states with two or more Higgs bosons. This approach is somewhat limited, though, on two accounts. Firstly, these subchannels cannot lead to electrically charged final states. Hence, they necessarily miss out on some couplings involving a charged Higgs boson, in parameter space regions of the 2HDMs where the neutral final state production processes these couplings might alternatively enter are kinematically unavailable. Secondly, there could exist further production channels (for neutral final states) offering access to many other triple-Higgs couplings, also needed to reconstruct the full EWSB potential. 

In this paper, we have therefore concentrated on EW-induced channels, where the initial state constitutes (primarily) of valence quark flavours, which annihilate via both electrically neutral and charged currents into neutral and charged 2-Higgs (and up to 3-Higgs) final states. We have shown that the production cross sections for several charged final states (precluded to the $gg$ and $b\bar b$ production modes) are large enough to be potentially accessible at the LHC, either during the Runs 2 and 3 or at its High Luminosity (HL-LHC) stage (depending on the parameter space configuration). We have also illustrated that such EW-induced channels can often be competitive with, when not overtaking, those induced by $gg$ and $b\bar b$ fusion, other than offering more probes of various triple-Higgs couplings. Finally, as these EW channels are often mediated by weak gauge bosons (i.e., $W^\pm$ and $Z$ states), they can provide sensitivity to couplings involving one of these and two Higgs bosons.

We have come to these conclusions after studying, as a preliminary step of a long-term investigation that will eventually include a complete detector simulation, the fully inclusive parton-level yield of the aforementioned EW channels. This study tackled the phenomenology of the so-called Type-I 2HDM, as illustrative for conditions which may emerge in other possible non-minimal Higgs constructs, in the presence of standard theoretical constraints as well as the latest experimental limits coming from EW precision data, collider searches for the Higgs boson(s), and measurements of the heavy flavour observables. 

In short, we advocate, alongside the time-honoured analyses based on QCD-induced processes, investigations of EW processes as well, which we have shown to offer improved and expanded sensitivity to both Higgs and gauge-Higgs structure of the underlying EWSB dynamics, which may or may not be the same as the SM one.

\section*{Acknowledgements} 

SMo is supported in part through the NExT Institute and the STFC Consolidated Grant ST/L000296/1. RE, WK and SMo are partially supported by the H2020-MSCA-RISE-2014 grant no.\ 645722 (NonMinimalHiggs). 



\providecommand{\href}[2]{#2}\begingroup\raggedright\endgroup

\end{document}